\documentclass[fleqn,usenatbib]{mnras}

\usepackage{newtxtext,newtxmath}

\usepackage[T1]{fontenc}
\usepackage{rotating}

\DeclareRobustCommand{\VAN}[3]{#2}
\let\VANthebibliography\thebibliography
\def\thebibliography{\DeclareRobustCommand{\VAN}[3]{##3}\VANthebibliography}

\usepackage{graphicx}	
\usepackage{amsmath}	

\title[Non-thermal radio components from the Orion]{Detection of non-thermal radio emission components from the Orion Nebula: stellar jets, cloud collision or feedback from stellar winds?}
\author[Rashid et al.]{
Md Rashid,$^{1,2}$\thanks{E-mail: rashidmd.astro@gmail.com}\thanks{Now at Tata Institute of Fundamental Research (TIFR), Mumbai}
Nirupam Roy,$^{2,7}$
Prasun Dutta,$^{3}$
Jagadheep D. Pandian,$^{4}$,
Sarita Vig,$^{4}$
Srijita Pal,$^{2}$ \newauthor
Arnab Chakraborty,$^{5}$ 
Samir Choudhuri$^{6}$
\\
$^{1}$Joint Astronomy Programme, Indian Institute of Science, Bangalore 560012, India\\
$^{2}$Department of Physics, Indian Institute of Science, Bangalore 560012, India\\
$^{3}$Department of Physics, IIT (BHU), Varanasi 221005, India\\
$^{4}$Department of Earth and Space Sciences, Indian Institute of Space Science and Technology, Trivandrum 695547, India\\
$^{5}$Department of Physics and McGill Space Institute, McGill University, Montreal, QC H31 2T8, Canada\\
$^{6}$Centre for Strings, Gravitation and Cosmology, Department of Physics, Indian Institute of Technology, Madras, Chennai 600036, India\\
$^{7}$Department of Physics, New Mexico Institute of Mining and Technology, Socorro, NM 87801, USA
}

\date{Accepted XXX. Received YYY; in original form ZZZ}

\pubyear{\the\year{}}

\begin{document}
\label{firstpage}
\pagerange{\pageref{firstpage}--\pageref{lastpage}}
\maketitle

\begin{abstract}
The Orion Nebula is the closest high-mass star-forming region, making it an ideal laboratory to investigate physical processes in complex star-forming environments. At radio frequencies, the dominant emission mechanisms are thermal bremsstrahlung and non-thermal synchrotron. H~{\sc ii} regions typically emit thermal radiation tracing the ionised gas; however, detecting and characterising non-thermal emission can provide insights into magnetic fields and the energy distribution of relativistic particles in star-forming regions. We have utilised the upgraded Giant Metrewave Radio Telescope (uGMRT) to study radio emission in the Extended Orion Nebula (EON) region. We present results from wide-band interferometric observations using uGMRT bands~3 and~4, probing a frequency range not covered by other sensitive radio interferometers. We produced deep continuum images with RMS noise levels of $\sim400\,\mu$Jy~beam$^{-1}$ in band~3 and $\sim200\,\mu$Jy~beam$^{-1}$ in band~4. We further generated in-band and broad-band spectral index maps using these images. To establish the robustness of the spectral index measurements, we conducted a detailed analysis using simulated uGMRT data. From the continuum spectral index analysis, we report the unambiguous presence of non-thermal radio emission in the EON region. To investigate its plausible origin, we correlated our results with multiwavelength observations, identifying a strong association between non-thermal emission and outflows from young stellar objects, while also exploring alternative explanations. In future, reliable broad-band radio spectral index measurements, together with dedicated multiwavelength observations, will be invaluable for resolving the origin of non-thermal emission in the Orion Nebula and other star-forming regions.
\end{abstract}



\section{Introduction}
Stars are born within the coldest and densest parts of the molecular cloud in the interstellar medium (ISM). At the end of their lives, stars leave behind both compact and diffuse remnants, the latter reverting to the ISM. This is the so-called star-gas cycle \citep{2007IAUS..235..268P}. However, during the main sequence phase, they influence the environment by radiative and mechanical feedback. Massive stars radiate far ultraviolet (UV) photons and ionise the ISM around them. The major constituent of this ionised region is ionised hydrogen (H~{\sc ii}), and hence, these are called the H~{\sc ii} region\footnote{Helium and other metals also form subsequent layers of ionised shells in order of their increasing ionisation potentials}. Star clusters hosting massive young stars are commonly found near the photodissociation regions (PDRs), the interface of the H~{\sc ii} region and cold molecular clouds \citep{2003A&A...399.1135D, 2004ASPC..322..375D, 2013AJ....145...78K, 2015A&A...582A...1D}. The interface region has turbulent shocked gases, which trigger further star formation \citep{1977ApJ...214..725E}. The star clusters are typically deeply embedded within the dust, which absorbs UV radiation from young massive stars and re-radiates in the infrared bands \citep{2025arXiv250101613Z}. Moreover, strong stellar winds from massive stars in these clusters inject further energy, heating the gases to very high temperatures \citep[up to millions of Kelvin; ][]{2008Sci...319..309G}. The shock-heated regions are potential sites for Cosmic Ray (CR) acceleration, giving rise to non-thermal processes \citep{2025arXiv250112767P}. The interaction of CR with the ambient ISM gives rise to emission that can be traced at gamma-ray, X-ray and radio wavelengths \citep{2017PPCF...59a4002P}. Furthermore, the mechanical energy of the wind can inflate bubbles which contain shocked swept-up mass from the ambient gas \citep{2006ApJ...649..759C, 2020A&A...639A...2P}.  In addition, in active star-forming regions, molecular outflows and jets from Young Stellar Objects (YSOs) introduce turbulence in the region and create shock fronts at different scales \citep{1993Natur.363...54A}. The bubbles and outflows sometimes interact with each other and create small-scale pockets of shocked gas \citep{2022A&A...663A.117K}. It is, hence, instructive to conduct detailed observations to study the complex interplay between different components to understand how stellar feedback regulates star formation in galaxies.%
\par
Sub-GHz continuum studies of star-forming regions offer key insights into the physical processes governing the interplay between star formation and the physical environment. Since non-thermal synchrotron emission becomes more prominent at lower frequencies, sub-GHz studies help probe non-thermal processes and distinguish them from thermal free-free emission from the ionised gas. This frequency regime is sensitive to diffuse, optically thick and thin regions, providing critical information about features like shocks, winds, and supernova remnants \citep{2023A&A...678A..72D}. With the advent of modern radio interferometers equipped with wide-band observation capabilities, it is now possible to probe these non-thermal emitting regions at low frequencies with unprecedented sensitivity and resolution. For example, the wide-band upgrade of the Giant Metrewave Radio Telescope \citep[uGMRT;][]{gupta17} offers an instantaneous bandwidth of up to 400 MHz at frequency ranges that are not currently covered by any other radio interferometer worldwide. These sensitive observations have led to the detection of non-thermal radio emission from star-forming regions \citep[e.g.][]{2016AJ....152..146N,2016MNRAS.456.2425V}. These pioneering works are important; however, as non-thermal emission in these regions is intrinsically faint compared to the thermal emission, establishing the validity of such detections is observationally challenging. In addition, given the complex environment of star-forming regions, extensive multi-wavelength investigations are required to understand the ambient physical scenarios which possibly give rise to such emission \citep{2019A&A...630A..73M,2019MNRAS.482.4630V}. Therefore, robust quantification of uncertainties is necessary to reinforce these detections, while multi-wavelength investigations remain pivotal in uncovering the underlying physical mechanisms.

\par The Orion Nebular region (M42) is the closest bright high-mass star-forming region in the Galaxy \citep[distance$\sim 417$ pc;][]{2007A&A...474..515M}. It is an ideal target to study the physical environment of star-forming regions as it hosts a diverse class of astrophysical objects such as YSOs \citep{2005ApJS..160..319G}, massive O-, B-, and A-type stars \citep{2008hsf1.book..483M}, multiple star clusters \citep{2004ApJ...605L..57G, 1997AJ....113.1733H}, H~{\sc ii} region \citep{2001PASP..113...41F}, ionisation front (IF) \citep{2009ApJ...693..285P}, Herbig–Haro (HH) objects \citep{2000AJ....119.2919B} and the layers of ionized and neutral gases collectively called Orion's veil \citep{2001ARA&A..39...99O}. The FUV radiation from the star $\theta^1$ Ori~\textit{C} \citep[spectral class O5.5; ][]{2007A&A...466..649K, 2009A&A...497..195K} is primarily responsible for ionising its H~{\sc ii} region around it and forming a spectacular IF known as Orion bar. The interactions between these components give rise to a complex and rich environment \citep{2009AJ....137..367O, 2020ApJ...891...46O}. Hence, it serves as a reference for our understanding of star formation and evolution processes, probing the physical environment and evaluating the reliability of astronomical methods. As a result, it is one of the most extensively studied sources in the Galaxy. 
\par
The radio continuum morphology of the Orion nebula has been studied in great detail in previous studies \citep{1990ApJ...361L..19Y, 1992MNRAS.254..291S, 1993A&AS...98..137F, 2001AJ....121..399S}. Previously, \citet{2001AJ....121..399S} have reported the detection of only thermal emission towards the Orion nebula using VLA data at 333~MHz and 1.5~GHz frequencies in their studies. The determined spectral index was $1.6\pm0.1$ at the optically thick central bright region, and away from the central region, the spectral index was found to be $0.1\pm0.1$. These values are consistent with theoretical expectations (will be discussed subsequently) since H~{\sc ii} regions are dominated by thermal free-free emission from the ionised gases peaking at low radio frequencies.  However, their study was carried out with very coarse angular resolution before the advent of wide-feed radio interferometers. As discussed, detecting non-thermal emission from star-forming regions is picking up in recent studies; revisiting the Orion nebular region with wide-band observations can provide new insight regarding the emission mechanisms and physical environment. Hence, in this work, we leverage the unique capabilities of uGMRT to study low-frequency emission from the Extended Orion Nebula \citep[EON; ][]{2008Sci...319..309G}. 
\par
The flux density ($S_\mathrm{\nu}$) of radio-emitting sources often scales with frequency ($\nu$) as a power law:
\begin{equation}
S_\mathrm{\nu} \propto \nu^{\alpha},
\end{equation}
where the exponent $\alpha$, known as the spectral index, helps distinguish emission mechanisms. For thermal bremsstrahlung, the spectral index varies with optical depth. Under optically thick conditions (optical depth, $\tau \gg 1$), the continuum spectrum steepens sharply, with $\alpha \approx +2$ \citep{rlch5}. As the material transitions to an optically thin state ($\tau \ll 1$), the spectrum flattens significantly, approaching $\alpha \sim -0.1$. In contrast, non-thermal synchrotron radiation exhibits steeper indices, typically spanning $-0.1 \geq \alpha \geq -2$ \citep{rlch6}. However, to account for measurement uncertainties in observations, values of $\alpha \leq -0.4$ are generally required to confidently classify emission as non-thermal \citep{1999ApJ...527..154K}.
\par
One of the primary advantages of wide-band interferometric observations is the ability to measure the spectral index of continuum emission in a spatially resolved manner. However, for low signal-to-noise ratio (S/N) cases, measurement of reliable spectral indices becomes challenging. Moreover, for extended sources with complex morphology, factors like wide-field imaging effects, \textit{uv}-coverage, the angular extent of the source, the frequency dependence of the primary beam structure, fractional bandwidth, and the spectral response of the telescope over the large fractional bandwidth in question add to the challenge. The \textit{uv}-coverage plays particularly a critical role. Since different angular scales are sampled by different baseline lengths, the distribution of \textit{uv}-points directly determines the spatial frequencies to which the observation is sensitive. Consequently, when measuring broad-band spectral index, a mismatch in \textit{uv}-coverage between observing bands can lead to inconsistencies in the recovered flux densities and, hence, in the derived spectral index. To minimise such effects, the higher-frequency image is typically convolved (or “downgraded”) to match the angular resolution of the lower-frequency image, ensuring that both images sample comparable spatial scales. Additionally, restricting the imaging to a common \textit{uv}-range across both bands may improve consistency. However, achieving identical \textit{uv}-distributions at different frequencies is practically impossible for any interferometric array, as the projected baselines scale inversely with wavelength.

\par
In principle, existing algorithms like MT-MFS can measure the in-band spectral index using instantaneous wide-band data \citep{rau2011}. However, \cite{2024ApJ...971...39R} showed that even for point sources, the MT-MFS algorithm could not accurately measure the spectral indices for S/N less than 100. One has to resort to the broad-band (also commonly referred to as inter-band) spectral index for more reliable measurement in low S/N scenarios, where the measurements are reliable at an S/N of 15 and above. However, since the study of \cite{2024ApJ...971...39R} was limited to point sources, the reliability of spectral index measurements as a function of S/N for extended sources is unclear. Errors introduced in the spectral index measurement due to wide-field imaging must be investigated before drawing any conclusions from observations.    

\par    
In this paper, we present low-frequency uGMRT observations in band 3 (300-- 500 MHz) and band 3 (635-- 735 MHz) towards the Orion nebular region. The remainder of the paper is organised as follows: In \S \ref{sec: observaton}, we have described the calibration and imaging of observational data from uGMRT; in \S \ref{sec: results}, we show results from the observations of uGMRT; in \S \ref{sec: simulation}, we describe the simulations of radio observations for an extended source and the reliability of determining the spectral index from such observations; in \S \ref{sec: discussion}, we discuss the implications of the observations and simulations along with the different possibilities that can give rise to the detected non-thermal emission and in \S \ref{sec: conclusion}, we summarise the study with a brief conclusion.

\section{Observations and data reduction}
\label{sec: observaton}
\subsection{Data}
\begin{table}
	\centering
	\caption{Observational details of uGMRT data for the Orion Nebular region.}
	\label{tab:obs_parm}
	\begin{tabular}{l|cc} 
		\hline
        Parameters & Band 3 & Band 4 \\
        \hline
        \hline
        Obs. proposal ID & 31\_106 & 36\_014 \\
		Obs. date & 9$^\text{th}$ Jan 2017 & 15$^\text{th}$ \& 16$^\text{th}$ Jun 2019 \\
		No. of antenna & 16 & 30 \\
		Ref. frequency (MHz) & 400 & 685 \\
		Channel width, $\delta\nu$ (kHz) & 12.2 &6.1\\
		Bandwidth, $\Delta\nu$ (MHz) & 200 & 100\\
		No. of channels & 1648 & 1648\\
		On-source time, $\tau$ (Hr) & 6.5 (approx.)  & 7.33 (approx.)\\
		\hline
	\end{tabular}
\end{table}
We have used uGMRT archival band 3 and band 3 observations of the Orion Nebular region. The band 3 data were observed on 09 January 2017 under proposal ID 31\_106 with 16 wide-feed antennae. The phase centre of the observation for this observation was $05^{\rm h}35^{\rm m}17\rlap{.}^{\rm s}5$, $-05^\circ23'37''$. The bandwidth (BW) observation was 200~MHz centred at 400~MHz. The primary calibrator was 3C147, and the gain calibrators were 0503+020 and 0607-085. The band 3 data was observed on 15 and 16 June 2019 under proposal ID 36\_014 with 30 wide feed antennae. The phase centre of this observation was $05^{\rm h}34^{\rm m}59\rlap{.}^{\rm s}8$, $-05^\circ25'09''$. The BW was 100~MHz in this observation, centred at 685~MHz. The flux calibrator was 3C147, and 0501-019 was used as the phase calibrator. The observational details are listed in Table~\ref{tab:obs_parm}. 
\subsection{Calibration}
We have used the Common Astronomical Software Application \citep[CASA;][]{2007ASPC..376..127M}\footnote{\url{https://casa.nrao.edu/}} version 5.6 for data monitoring, reduction, calibration, and further processing including imaging. The calibrations were done using a script running standard CASA tasks. The general steps\footnote{The parameters and other details for the steps for all datasets can be found here: \url{https://github.com/rashid-astro/orion-calibration.git}} are as follows:
\begin{itemize}
    \item Initial flagging \& calibration: Some routing flagging of bad data points such as shadow, zeroes clipping, quack mode, end channels and automated flagging in `tfcrop' and `rflag' mode with mild threshold using task \texttt{flagdata}  prior to any calibration. This was followed by delay and band-pass calibration using task \texttt{gaincal} and \texttt{bandpass}.
    \item Flagging:  Automated RFI flagging of the corrected calibrator data in `tfcrop' and `rflag' mode using \texttt{flagdata} task.
    \item Iterative calibration: Repeating the flagging and calibration iteratively to remove persistent RFIs and improve calibration solutions.
    \item Final gain solutions: Final gain solutions were determined using the \texttt{gaincal} task and applied to the target data using the \texttt{applycal} task.
    \item Target flagging: Automated target flagging using `tfcrop' and `rflag' in \texttt{flagdata}. Finally, examination of the target data and manual flagging of any leftover RFI after automated flagging.
\end{itemize}
There is a known issue\footnote{\url{http://www.ncra.tifr.res.in/ncra/gmrt/gmrt-users/recent-gmrt-updates}} of flux density amplitude offset in the calibrated visibilities in the baselines between GMRT central square antennas with respect to baselines from other antennas due to grounding issues in the backend receiver system in the GMRT Wideband Backend (GWB) data, observed between October 2018 and 3 December 2020. The band 4 data for both the dates (15$^\text{th}$ and 16$^\text{th}$ June 2019) had a slight offset ($\sim3$\%) due to this issue. We tackled this problem by following the calibration strategy prescribed by the observatory. For band 4 datasets, the data were split into two: one containing baselines between central square antennas and the other containing the baselibaselines from the remaining All the datasets were then calibrated separately. After obtaining satisfactory calibration solutions, we applied them to the respective datasets and concatenated them before further analysis. 

\subsection{Self-calibration \& imaging}
\begin{table}
	\centering
	\caption{Imaging parameters for the \texttt{tclean} task.}
	\label{tab:image_parm}
	\begin{tabular}{lr} 
		\hline
       
        Parameters & values\\
        \hline
         \hline
		imsize  & [2048, 2048]\\
		  cell ($''$)& 2 \\ 
	    specmode & mfs \\
        gridder & widefield \\
        wprojplane & 16 \\
        deconvolver & MTMFS \\
        nterm & 2 \\
        scales & [$0''$, $10''$, $50''$, $250''$] \\
        weighting & Briggs \\
        robust & 0 \\
        usemask & auto-multithresh \\
	\hline
	\end{tabular}
\end{table}
Self-calibration was done with continuum model images to further improve the calibration of the target data. For this, we adopt the following strategy. First, a relatively shallow CLEANing was done to get an initial model image of the targets. The Fourier transform of the model was saved as a model data column during imaging. The gain solutions are then obtained by optimising between the initially calibrated and model data columns. These solutions are examined and applied to the data. Following this, we ran tfcrop on the residual data to get rid of any low-level RFI in the residuals. This process is repeated a few times until the noise in the image saturates. We have done five rounds of phase-only self-calibration and one round of full gain (amplitude and phase) self-calibration on band 3 data. Similarly, seven rounds of phase-only self-calibration and one round of full gain self-calibration were done on band 4 data before obtaining the final continuum image of the Orion Nebular region.
\par
The \texttt{tclean} task was used for continuum imaging with ``wide-field" gridder and Multi Scale-Multi Term Multi Frequency Synthesis \citep[MS-MTMFS; ][]{rau2011} deconvolver. Table~\ref{tab:image_parm} lists the important imaging parameters. We have used Briggs's weighting scheme \citep{briggs95} with `robust' set to 0 to have an optimal balance between noise and resolution. \texttt{Auto-multithresh} masking was employed to automatically mask emission for deconvolution during CLEAN, based on appropriately chosen thresholds for each band. The number of iterations was set to an arbitrarily large number. Finally, we did the primary beam correction to the images using the uGMRT-specific task \texttt{wbpbgmrt} \citep{2021ExA....51...95K}\footnote{https://github.com/ruta-k/uGMRTprimarybeam}, which is equivalent to the CASA task ``widebandpbcor''.

\section{Observational Results}
\label{sec: results}
\begin{figure}
    \centering
     \includegraphics[trim={1cm 3cm 3cm 4.5cm},clip,width=1\linewidth]{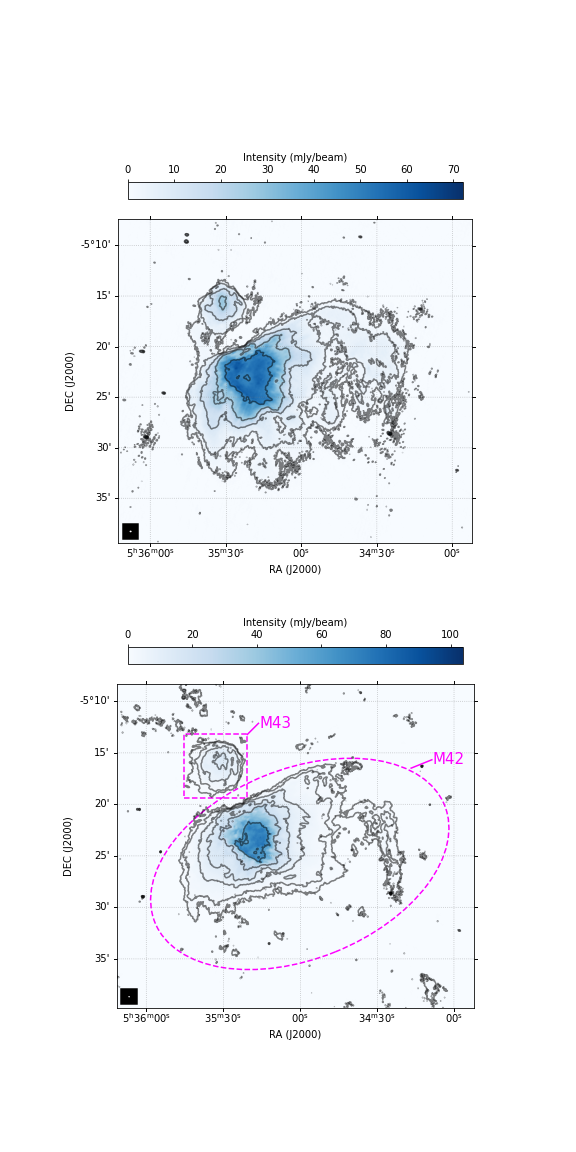}
    \caption{Continuum image centred at 400 MHz from band 3 \textbf{(top)},and at 685 MHz from band 4 \textbf{(bottom)} of uGMRT. The contour levels are at [1,3,10,20,40] $\times 3 \sigma $ for band 3 image ($\sigma_{\mathrm{B3}}=400\mu$Jy/beam), and at [1,3,10,20,40,100] $\times 3 \sigma$ for band 4 image ($\sigma_{\mathrm{B4}}=200\mu$Jy/beam). The synthesised beams (ellipse) are shown in the bottom left corner of the respective image.}
    \label{fig:cont-b3-4}
\end{figure}

\begin{figure*}
    \centering
    \includegraphics[width=0.98\linewidth]{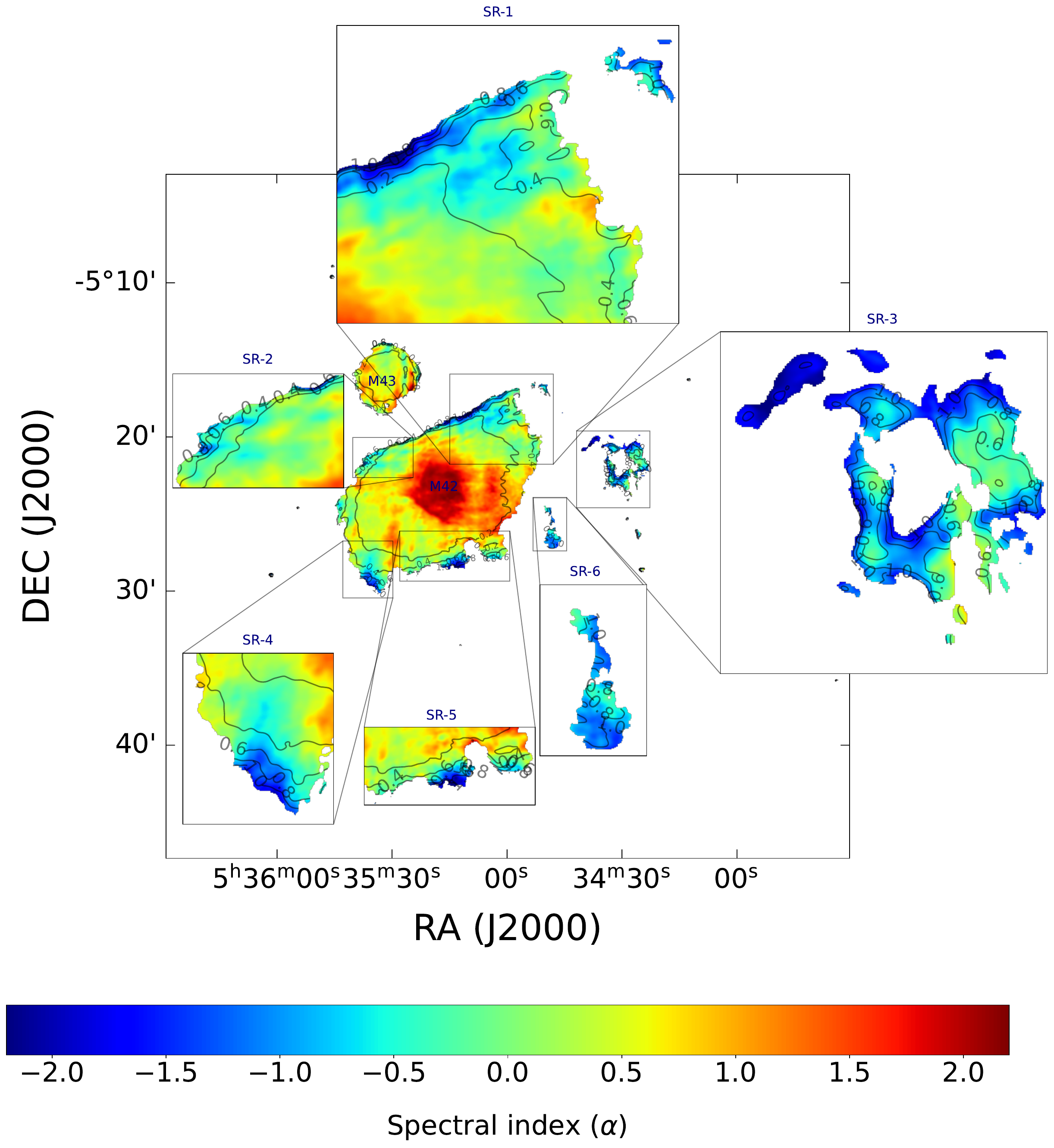}
    \caption{Broad-band spectral index map of the Orion H~{\sc ii} region obtained from 400 MHz and 685 MHz images of uGMRT. The overlaid labelled contours depict the uncertainty in the spectral index, ranging from $0.2<\sigma_{\alpha}<1$. The regions with negative spectral indices have been magnified.}
    \label{fig:spectra-index-orion}
\end{figure*}
We have obtained high-resolution deep continuum images of the Orion Nebula with root mean square (RMS) noise of 400 $\mu$Jy-beam$^{-1}$ in band 3 and 200 $\mu$Jy-beam$^{-1}$ in band 4. The synthesised beam sizes of the images are $10''\times 6''$ and $5.6''\times 4.6''$, respectively. Figure~\ref{fig:cont-b3-4} shows the continuum image overlaid with contours at [1,3,10,20,40] $\times 3 \sigma$ for band 3 image, and at [1,3,10,20,40,100] $\times 3 \sigma$ for band 4. This is the deepest interferometric continuum image availabl at sub-GHz frequencies with such a high angular resolution. The images feature two well-known objects: (i) Orion nebula or M42, a complex and irregularly shaped H~{\sc ii} region in the central part of the image; (ii) A spherically shaped  H~{\sc ii} region M43 to the northeast of M42, ionised by a B type star HD 37061 (NU Ori) \citep{2011A&A...530A..57S}. In this study, we will focus on the M42 region and for the rest of the discussion, we will refer to this region only.

\par
In addition to the continuum images, our study produced an in-band spectral index map during MTMFS deconvolution using two Taylor terms. However, the fractional bandwidth ($\Delta\nu/\nu_0 \times100$) of band 4 observation was only $\sim 15\%$, making the in-band spectral index map practically of little to no use. On the other hand, the band 3 data had a fractional bandwidth of $50\%$, so it produced somewhat useful spectral index values compared to the band 4 in-band spectral index map. The band 3 in-band spectral index map is shown in Appendix \ref{app:a}. While the map shows positive spectral indices indicative of thermal free-free emission in the central part of M42, there are several regions that manifest negative spectral indices that are characteristic of non-thermal emission. However, as mentioned earlier, such in-band spectral index maps are not reliable enough to be used for scientific conclusions, except in regions of very high S/N.

\par
We have made a ``two-band" or broad-band spectral index map of the Orion Nebula using band 3 ($\nu_1=400$~MHz) and band 4 ($\nu_2=685$~MHz) images. We first convolved both the images to a common resolution of $10.5'' \times 7.5''$, which encompasses the synthesised beam of both the images from two bands. Then we regridded them to a common pixel size of $1.3''$. These steps help mitigate spurious estimation of the spectral index due to resolution and sampling mismatches. Then, we put a threshold of 15$\sigma$ \citep[based on constraints from ][]{2024ApJ...971...39R} to each image so as to include only pixels with high S/N and estimated the spectral index pixel-by-pixel using the formula:
\begin{equation}
\label{eq:bbsp}
\alpha = \frac{\log\left( \frac{S_{685}}{S_{400}} \right)}{\log\left( \frac{685}{400} \right)},
\end{equation}
to the overlapping valid pixels of the two images. We also calculated the uncertainty in the spectral index on a per-pixel basis using error propagation \citep[see \S 2.2 of][]{2024ApJ...971...39R}. Figure~\ref{fig:spectra-index-orion} shows the broad-band spectral index map of the Orion Nebular region. The overlaid contours indicate the uncertainty, with the values labelled on individual contours. The central part of M42 mostly shows $\alpha=+2$, which is consistent with optically thick thermal emission. Beyond the bright Huygens region \citep{2013ApJ...762..101V}, $\alpha$ gradually drops to zero, indicative of optically thin thermal emission. Further away from the bright central part, multiple (shocked) regions (magnified as SR-1, SR-2, SR-3, SR-4, SR-5,  and SR-6 in Figure~\ref{fig:spectra-index-orion}) consist of a complex mix of positive, zero, and negative $\alpha$. Although most of the negative $\alpha$ values belong to higher uncertainty regions ($0.2<\sigma_{\alpha}<1$), there are also regions with low uncertainty ($\sigma_\alpha < 0.2$). We have shown the distribution of spectral index with respect to S/N in the Appendix \ref{app:alpha-SN}. 

\par
The Orion Nebula is an H{\sc ii} region dominated by emission from ionised gas, and thus a spectral index characteristic of thermal free–free emission is expected and has been reported in previous radio studies. However, the detection of a negative spectral index, indicative of a non-thermal component, is a new finding that has not been reported in earlier radio studies of this region. Since this interpretation relies on the spectral index map, it is essential to establish the reliability of the measured spectral index, particularly in the negative regime. The broad-band spectral index map of Figure~\ref{fig:spectra-index-orion} has been obtained by applying S/N constraints obtained from the simulations of unresolved sources reported in \cite{2024ApJ...971...39R}.  However, the extended emission with varying S/N across an extended source structure may be affected by other systematics of wide-field imaging. We use simulated observations of the extended source to verify the reliability of spectral index estimation.

\section{Simulation}
\label{sec: simulation}
 
\begin{figure}
    \centering
    \includegraphics[width=1.0\linewidth]{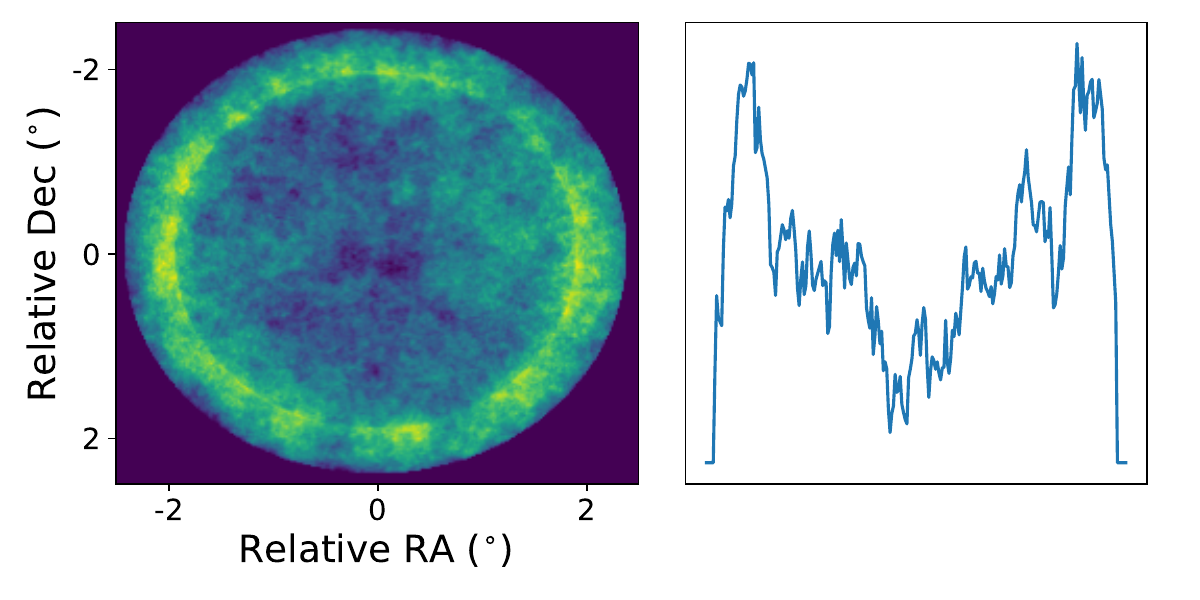}
    \caption{2D Spherical shell model \textbf{(left)} for extended source simulation. A profile of brightness along one of the diameters is shown \textbf{(right)}.}
    \label{fig: model}
\end{figure}

\begin{figure}
    \centering
    \includegraphics[trim={0.3cm 0cm 1.5cm 1cm},clip,width=0.99\linewidth]{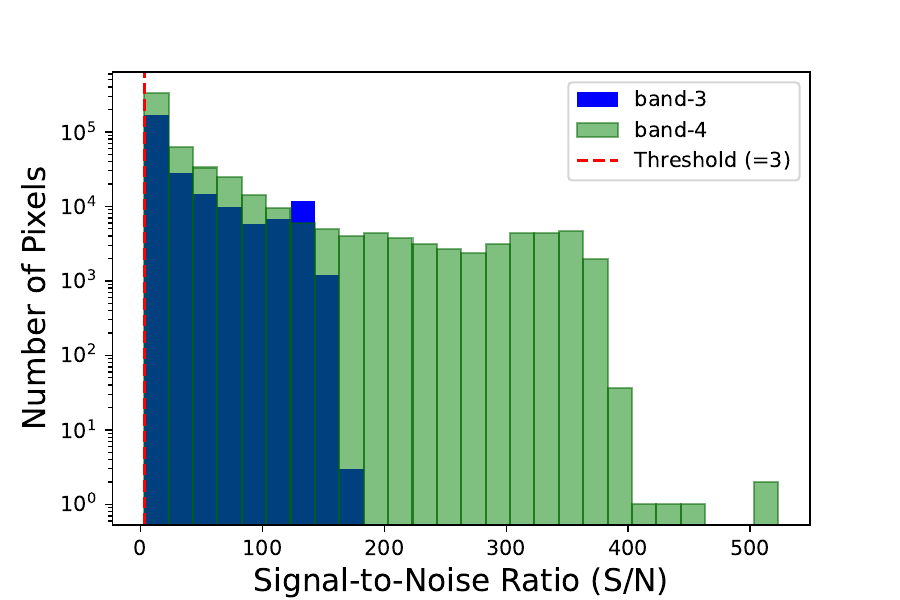}
    \caption{Histogram of S/N for band 3 (\textbf{blue}) and band 4 (\textbf{green}) continuum image. A cutoff at S/N=3 is placed in the plot, denoted by a dashed-red vertical line. }
    \label{fig: snhist}
\end{figure}

\subsection{Model}
The spectral index measured from wide-bandwidth interferometric observations can be highly influenced by the extended source geometry. This is because an interferometer does not measure
total flux directly, but rather samples the spatial Fourier components of the sky brightness distribution through its \textit{uv}-coverage. Different angular scales correspond to different regions of the \textit{uv}-plane, and the completeness of these samples depends both on the array configuration and the observing frequency \citep{1992ARA&A..30..575C, 2017A&A...598A..78I}. For a source with complex or asymmetric morphology, variations in spatial frequency sampling between two frequencies can lead to unequal recovery of diffuse emission, which in turn biases the spectral index measurement. If large-scale emission is partially resolved out at one frequency but not the other, the resulting spectral index will appear artificially steep or flat, depending on which
spatial components are preferentially missing. Therefore, reliable spectral index estimation for extended sources requires careful matching of angular resolution and \textit{uv}-range between frequency bands \citep{rau2011}. However, as noted earlier, a perfectly matched \textit{uv}-coverage cannot be achieved. We therefore simulate observations of a model with a known spectral index to assess how accurately the spectral index can be recovered for a fixed configuration of uGMRT.

In this study, we are particularly interested in measuring the spectral properties of the non-thermal emission from shocks. To simulate this geometry of a typical shocked emission region, we choose the shape of a spherical shell. We have generated a 2D projected model of an optically thin spherical shell. The model source spans $5'$ in angular extent and is embedded within an image field of view (FoV) of $\approx 41'$, similar to the band 4 primary beam of the uGMRT. A zoomed-in version of the model (left), along with the profile of the brightness along a diameter (right), is shown in Figure~\ref{fig: model}. We have added fluctuations to the smooth brightness profile to mimic an astrophysical source. The smooth brightness profile can be given as:
\begin{equation}
\begin{split}
S_\mathrm{{smooth}} & = 2 S_0\left( \sqrt{R_2^2 -R^2}-\sqrt{R_1^2 - R^2}\right) \quad & \text{for} \quad R \leq R_1 \\
                    & = 2 S_0\left( \sqrt{R_2^2 -R^2} \right) \quad & \text{for} \quad R_2 \geq R \geq R_1
\end{split}
\end{equation}
where S$_0$ is constant amplitude, set to $1\mu$Jy/pixel, R$_1$ and R$_2$ are the inner and outer radii of the shell, and R is the arbitrary radius. We have taken the shell thickness ($R_2 -R_1$) to be 20\% of the source extent. For the fluctuating component, we have followed the prescription of \cite{2014MNRAS.445.4351C} and generated the diffuse synchrotron emission model, taken from a statistically homogeneous and isotropic Gaussian random field whose properties are completely specified by the angular power spectrum given by: 
\begin{equation}
    C_l^M=A \times \left(\frac{1000}{l}\right)^\beta,
\end{equation}
where $l$ is the angular multipole, $M$ is the azimuthal index, $A = 513$ mK$^2$ and $\beta$ = 2.34. First, the Fourier
components of the brightness temperature fluctuations ($\delta$T) are generated using the relation
\begin{equation}
    \delta T=\sqrt{\frac{\Omega C_l^M}{2}}[x(U) +iy(U)],
\end{equation}  
here, $\Omega$ is the total solid angle of the simulation, and $x(U)$ and
$y(U)$ are independent Gaussian random variables with zero mean
and unit variance. Fourier transformation was used to generate the corresponding specific intensity fluctuations in all pixels of size $1.2'' \times 1.2''$.

\par Next, we have used this 2D model image to generate a spectral cube to be used as our final model wherein the intensity is scaled as a power law in frequency  $I\sim\nu^\alpha$, where $\alpha$ is the spectral index. We have chosen a fiducial value of $\alpha =-0.7$, which is roughly the median spectral index of our observed spectral index map of the Orion Nebula, and it is also the average spectral index of the radio emission from star-forming galaxies at $\approx 1$~GHz \citep{1982A&A...116..164G, 2018A&A...611A..55K}.
\begin{figure}
    \centering
    \includegraphics[width=1\linewidth]{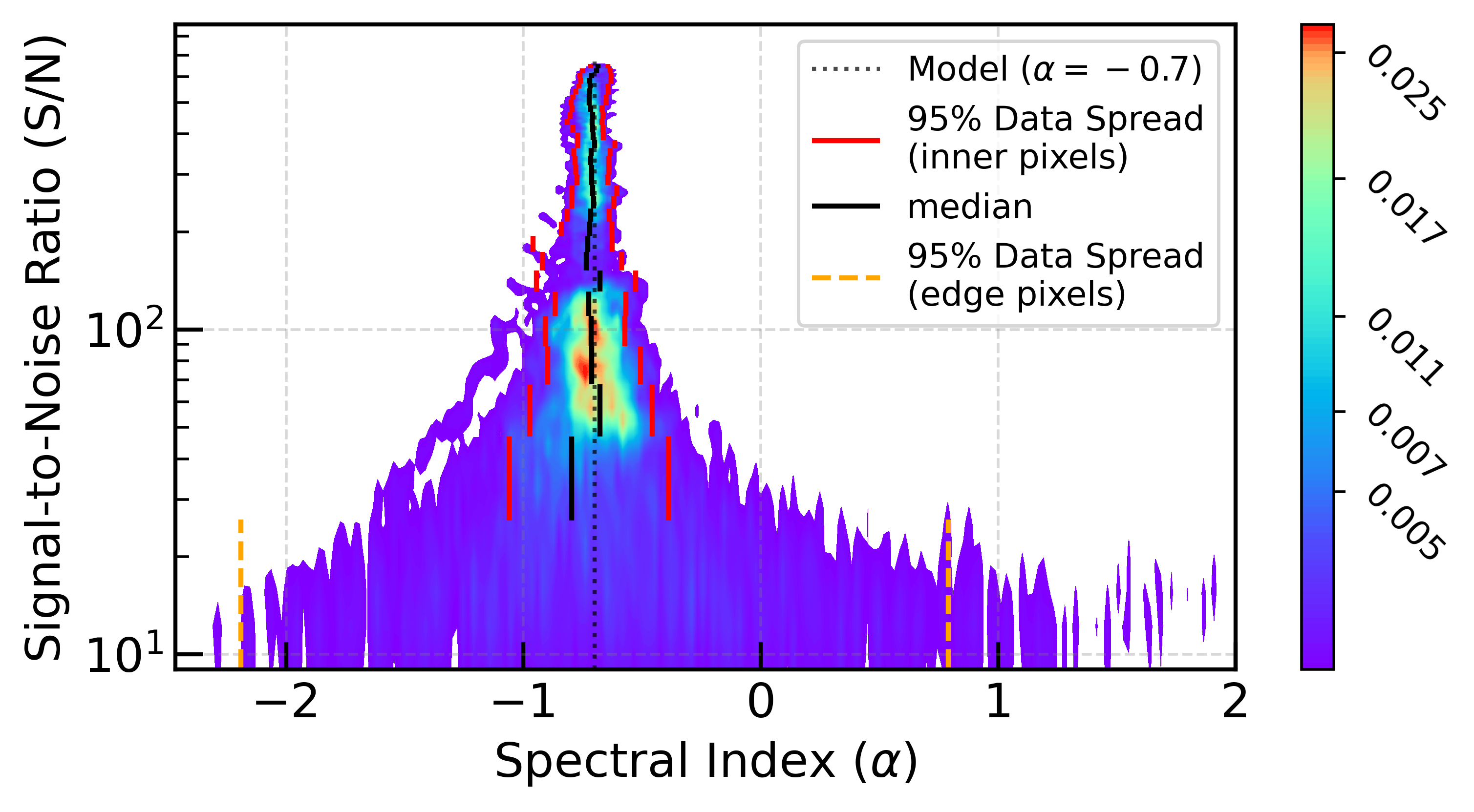}
    \caption{Two-dimensional kernel density estimate (KDE) of signal-to-noise ratio (S/N) vs. retrieved spectral index ($\alpha$) for simulated spectral index map. The fiducial (model) spectral index value for the simulation is $\alpha_{\rm{true}}=-0.7$, indicated by the dashed black line. The colour map represents normalised data point density. Solid red bars denote the 95\% data spread across thirty bins of S/N, and the black bar represents the median. The plot comprises data from $\sim38000$ pixels of the spectral index map from the simulated uGMRT data. The S/N axis is displayed on a logarithmic scale.}
    \label{fig:diff-sim-snr-sp}
\end{figure}
\subsection{Simulating data \& imaging}

Using the model spectral cube, we simulated the visibility data in uGMRT configuration by taking the Fourier transformation. We have used a C-based
code, \texttt{VISFITSGRID}, developed within the collaboration. The procedure begins by reading a
model FITS image cube (representing the diffuse or extended sky brightness distribution) and
a template \texttt{UVFITS} file containing the actual \textit{uv}-tracks from a uGMRT observation. For each frequency channel, the two-dimensional Fourier transform of the sky brightness distribution
was computed using the \texttt{FFTW3} library to generate complex visibility components. The Fourier transformation was performed as a real-to-complex 2D Fast Fourier Transform (\texttt{FFT}), which efficiently computes the complex spatial frequency representation of the image. The resulting complex visibilities were then interpolated and gridded onto the nearest sampled \textit{uv}-coordinates from the template \texttt{UVFITS} file, thus reproducing the effect of the actual interferometer’s spatial frequency sampling. This gridding step ensures that the simulated visibilities experience the same incomplete \textit{uv}-coverage as in real observations, which is crucial for studying the reliability of spectral index recovery. The visibility data were then written back into the \texttt{UVFITS} structure, producing a synthetic dataset that could be processed through the same imaging workflow used for real observations.

\par
For the observational configuration, we have placed the source at the observation phase centre ($05^{\rm h}35^{\rm m}17\rlap{.}^{\rm s}5$, $-05^\circ23'37''$), which is taken the same as the coordinate of the centre of M42 to match the \textit{uv}-distribution of the observation. Furthermore, we have considered 30 antennae with 6 hours of on-source time to achieve a reasonable \textit{uv}-coverage. Band 3 data have been simulated for a frequency range of 300-- 500 MHz (bandwidth $=200$ MHz), and band 4 data have been simulated for 550-- 650 MHz (bandwidth $= 100$ MHz). All the simulation parameters were chosen motivated by our observational data to make the simulation more realistic. A random noise per visibility data point has been added to the dataset of both bands to cover appropriate S/N ranges in the image domain. Noise levels in the visibility domain are scaled such that the RMS noise values in the resulting images closely match the RMS noise levels obtained in the actual observational data. We have used the \texttt{setnoise} of the CASA simulator tool to add noise to the visibility data \citep{2024ApJ...971...39R}. 

\par
We have reached a peak S/N of $\sim200$ in band 3 and $\sim550$ in band 4 for the M42 region in the radio continuum image from observations. A histogram of S/N for pixels with S/N greater than 3 is shown in Figure~\ref{fig: snhist}. However, due to limited computational resources, we were unable to populate the desired S/N range with a sufficient number of pixels across the entire range within a single simulation. Hence, we conducted two simulations to cover the desired S/N range. In the first simulation, the S/N range of 5 to 150 was populated; in the other simulation, the range was 5 to 650. This strategy gave us the advantage of populating the desired S/N range with an adequate number of pixels throughout the range. As a result, the statistical inferences are more robust. 

\par
We have imaged the data using the MS-MTMFS deconvolver with two Taylor terms. We used three Gaussian kernels with scales of $10''$, $50''$, and $2.5'$  along with zero scale during the multiscale deconvolution of our extended source. We have used `Briggs' weighting with robust=0. The pixel size was set to $2''$ in both bands. The rest of the imaging parameters are maintained consistent with those used for the continuum imaging of the observational data.

\subsection{Results from continuum spectral analysis}
\begin{table*}

\centering
\caption{Percentile statistics of measured spectral indices for different S/N ranges.}
\label{tab:sn_percentiles}
\begin{tabular}{lcccccccc}
\hline
&&&&$\Delta\alpha_\text{lower}$&$\Delta\alpha_\text{upper}$&$\alpha_\text{bias}$\\ 
S/N Range & $\alpha_{2.5}$ & $\alpha_{50}$ & $\alpha_{97.5}$ & 
($\alpha_{50}-\alpha_{2.5}$) & ($\alpha_{50}-\alpha_{97.5}$) &
($\alpha_{\rm true}-\alpha_{50}$) \\
\hline
\hline

26--47 & -1.06 & -0.80 & -0.39 & 0.26 & -0.41 & 0.10 \\
47--68 & -0.97 & -0.68 & -0.46 & 0.29 & -0.22 & -0.02 \\
68--89 & -0.90 & -0.71 & -0.51 & 0.19 & -0.2 & 0.01 \\
89--110 & -0.91 & -0.72 & -0.57 & 0.19 & -0.15 & 0.02 \\
110--131 & -0.87 & -0.72 & -0.57 & 0.15 & -0.15 & 0.02 \\
131--152 & -0.95 & -0.68 & -0.53 & 0.27 & -0.15 & -0.02 \\
152--173 & -0.92 & -0.74 & -0.59 & 0.18 & -0.15 & 0.04 \\
173--194 & -0.96 & -0.73 & -0.63 & 0.23 & -0.10 & 0.03 \\
194--215 & -0.84 & -0.72 & -0.63 & 0.12 & -0.09 & 0.02 \\
215--236 & -0.82 & -0.72 & -0.64 & 0.10 & -0.08 & 0.02 \\
236--257 & -0.80 & -0.70 & -0.62 & 0.10 & -0.08 & 0.00 \\
257--278 & -0.80 & -0.71 & -0.61 & 0.09 & -0.10 & 0.01 \\
278--299 & -0.77 & -0.71 & -0.65 & 0.06 & -0.06 & 0.01 \\
299--320 & -0.78 & -0.71 & -0.64 & 0.07 & -0.07 & 0.01 \\
320--341 & -0.78 & -0.72 & -0.64 & 0.06 & -0.06 & 0.02 \\
341--362 & -0.79 & -0.71 & -0.63 & 0.08 & -0.08 & 0.01 \\
362--383 & -0.77 & -0.70 & -0.62 & 0.07 & -0.08 & 0.00 \\
383--404 & -0.77 & -0.71 & -0.66 & 0.06 & -0.05 & 0.01 \\
404--426 & -0.79 & -0.71 & -0.66 & 0.08 & -0.05 & 0.01 \\
426--447 & -0.82 & -0.71 & -0.67 & 0.11 & -0.04 & 0.01 \\
447--468 & -0.80 & -0.71 & -0.67 & 0.09 & -0.04 & 0.01 \\
468--489 & -0.79 & -0.72 & -0.67 & 0.07 & -0.05 & 0.02 \\
489--510 & -0.80 & -0.72 & -0.65 & 0.10 & -0.12 & 0.02 \\
510--531 & -0.79 & -0.72 & -0.65 & 0.07 & -0.07 & 0.02 \\
531--552 & -0.78 & -0.72 & -0.64 & 0.08 & -0.08 & 0.02 \\
552--573 & -0.77 & -0.72 & -0.64 & 0.07 & -0.08 & 0.02 \\
573--594 & -0.76 & -0.72 & -0.63 & 0.04 & -0.09 & 0.02 \\
594--615 & -0.76 & -0.71 & -0.63 & 0.05 & -0.08 & 0.01 \\
615--636 & -0.75 & -0.69 & -0.64 & 0.06 & -0.05 & -0.01 \\
636--657 & -0.72 & -0.68 & -0.64 & 0.04 & -0.04 & -0.02 \\

\hline
\end{tabular}
\end{table*}

From the image of two bands, we have obtained a broad-band spectral index map for the simulated data. First, we have smoothed the images to a common resolution of $20''$ (encompassing the synthesised beam of images from both bands) and regridded to the same pixel size. For each pixel with S/N$\geq5$ in both bands, we calculated the spectral index values on each overlapping pixel using equation \ref{eq:bbsp}. Figure~\ref{fig:diff-sim-snr-sp} shows the kernel density estimated (KDE) plot of the spectral index values (obtained from all valid pixels on the spectral index map) versus the S/N. KDE bandwidth was optimised using Scott's rule with an adjustment factor of 0.11. We also note a relative drop in the number density of points near S/N of 150. This is the overlapping range of two simulations, having individual low data density. However, we have checked the statistical consistency of these two data distributions and found that the Inter Quantile Range (IQR$_{13}$) follows a similar trend in this range.

\par
To quantify the reliability of the measured spectral index values, we divided the S/N range into 30 bins and computed the 2.5, 50, and 97.5 percentiles ($\alpha_{2.5}$, $\alpha_{50}$, and $\alpha_{97.5}$ from here on) for each bin. The interval between the $\alpha_{2.5}$ and $\alpha_{97.5}$ percentiles encloses 95\% of the data, while the $\alpha_{50}$ represents the median of the retrieved spectral index distribution. At the edge of the source, due to low S/N and sparse \textit{uv}- coverage (note that ideal FFT requires an infinite number of Fourier components), particularly at short spacings,
causes poor reconstruction at the boundary of the source, when using CLEAN-based algorithms
\citep{2008A&A...490..455C}. Hence, to minimise the influence of imaging artefacts that are commonly present near the source edges in estimating uncertainty statistics, we included only those pixels located at least one synthesised beam inside the source boundary (here, the boundary is chosen at S/N = 10). In Figure~\ref{fig:diff-sim-snr-sp}, the 95\% spread of each bin is shown by red bars, the bin median, $\alpha_{50}$ is indicated by solid black bars, and the dotted black line represents the fiducial value, $\alpha_{\text{true}}=-0.7$, which is adopted as the model spectral index in the simulation.

\par
For the lowest S/N range (26–47), $\alpha_{2.5}$, $\alpha_{50}$ and $\alpha_{97.5}$ in this range are –1.06, –0.8 and –0.39, respectively. The distribution exhibits a noticeable asymmetry in this bin, with a slight skew toward more negative values. Furthermore, we observe a mild deviation of the median from the model value ($\alpha_\text{bias}=\alpha_\text{true} -\alpha_{50}= 0.1$), which we define as the bias. Although minute asymmetries and bias are present in other bins as well, this bin shows the clearest deviation. For the next S/N range (48–67), the percentiles shift to –0.97, –0.68, and –0.46. As expected, the spread progressively narrows with increasing S/N. For the 68–89 range, the spread falls approximately below $\pm0.2$ around the model value and continues to decrease, except for a few bins where the density of data is a little sparse compared to other bins. The percentile statistics in each bin are listed in Table \ref{tab:sn_percentiles}.

\section{Discussion}
\label{sec: discussion}
In this study, we have mapped the radio emission from the EON at <1GHz frequencies. From continuum spectral index measurements, we confirm thermal free-free emission from H~{\sc ii} region powered by $\theta^1$~Ori~C. Intriguingly, we also found signatures of non-thermal radio emission from some regions of the EON. In addition, we have obtained constraints from simulation for reliable spectral index measurement from extended source geometry. In the subsequent discussion, we further establish the robustness of the detection of thermal and non-thermal emission. We will also discuss possible physical scenarios that may give rise to such emission by comparing our findings with the results from studies at other wavelengths. 

\subsection{Free–free emission from the Strömgren sphere}

\begin{figure}
    \centering
    \includegraphics[width=1.0\linewidth]{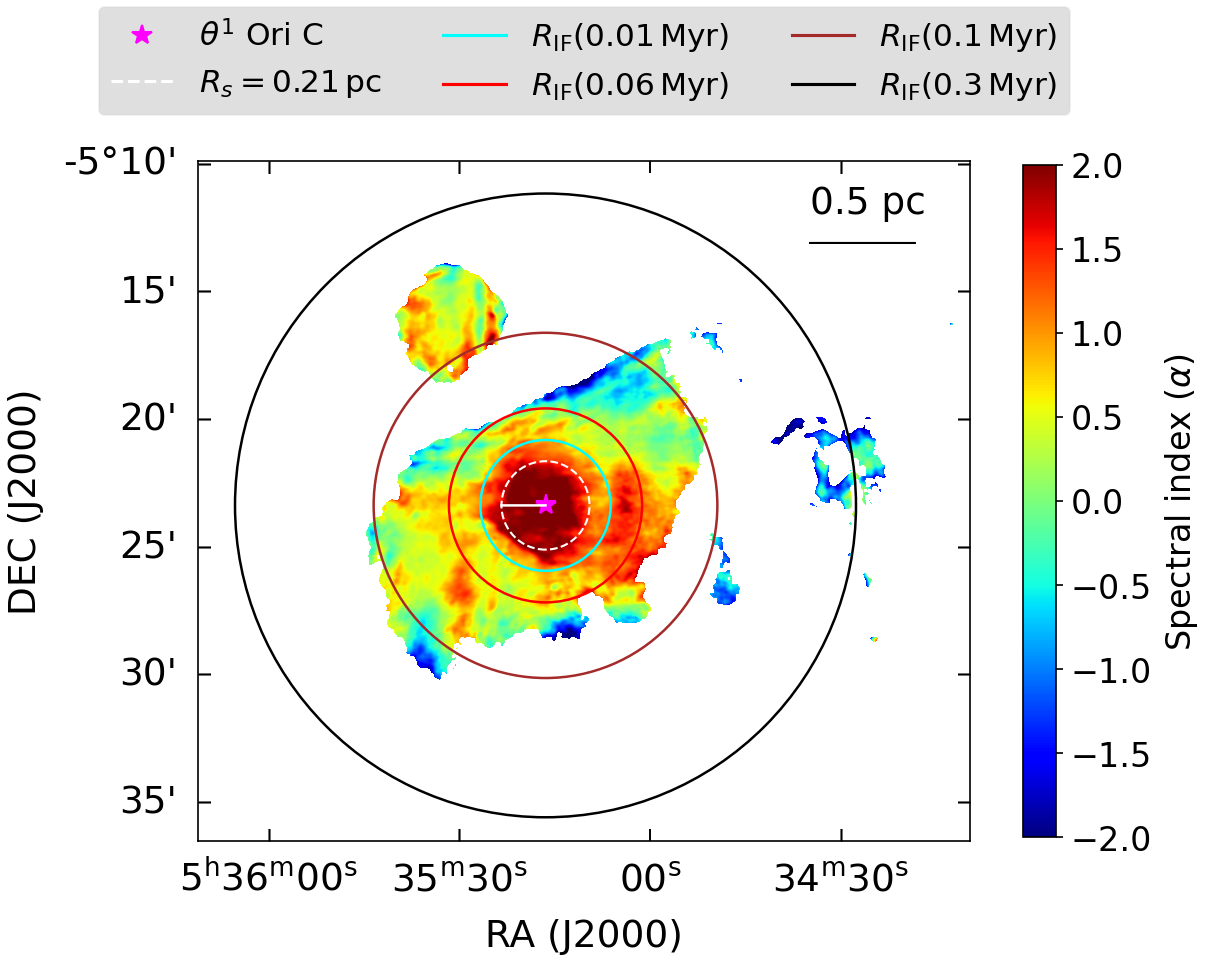}
    \caption{Estimated extent of Str\"{o}mgren sphere and ionisation front for different lifetimes of ionising star is overlaid on spectral index map. The position of the main ionising star $\theta^1$~Ori~C is shown with a magenta marker.}
    \label{fig:stromgren}
\end{figure}
The Str\"omgren radius of an H~{\sc ii} region is given by 
\begin{equation}
\label{stromgren}
R_{s} = \left( \frac{3Q_0}{4\pi\,\alpha_{B}\,n_{0}^{2}} \right)^{1/3},
\end{equation}
where $Q_0$ is the ionising photon emission rate, $\alpha_{B}$ is the case~B recombination coefficient, and $n_0$ is the number density. Assuming that the dominant ionizing source of the Orion Nebula, $\theta^1$~Ori~C, is an O7~V star \citep{simon2006}, we adopt $Q_0 \sim 10^{49}\,\mathrm{s}^{-1}$ \citep{martin2005} and $\alpha_{B} \sim 2.6 \times 10^{-13}\,\mathrm{cm^3\,s^{-1}}$ at $T_e \approx 10^4$~K \citep{storey1995}. The electron density in the Huygens region varies between $10^3$ and $10^4\,\mathrm{cm^{-3}}$; here we adopt $n_0 \sim 5700\,\mathrm{cm^{-3}}$, consistent with \citet{esteban1998}. Substituting these values into equation~\ref{stromgren} yields a Str\"omgren radius of $R_s \sim 0.21$~pc for the ionised region surrounding $\theta^1$~Ori~C. The dashed white circle at the central part of the Orion Nebula in Figure~\ref{fig:stromgren} illustrates the inferred Str\"omgren sphere overlaid on the spectral index map. The Huygens region of the Orion Nebula is known to exhibit high optical depth at sub-GHz radio frequencies \citep{odell2010,subrahmanyan1992}. 
The estimated Strömgren radius roughly matches the physical extent of the optically thick emission region with $\alpha\sim+2$, which is expected from ideal H~{\sc ii} regions dominated by free--free emission in the optically thick regime \citep{1999ApJ...527..154K}.

\par
However, the IF expands due to overpressure, and as a function of time, the expansion radius is approximated by:
\begin{equation}
\label{eq:rif}
    R_{IF} (t) = R_s \left( 1 + \frac{7}{4} \sqrt{\frac{4}{3}}\frac{c_s t}{R_s} \right)^{4/7},
\end{equation}
where $c_s$ is the sound velocity of in the medium \citep{2006ApJ...646..240H}. The age of the Orion Nebula is often quoted between $10^4-10^6$ yr, i.e., 0.01--1 Myr in the literature \citep{1963ApJ...138..294V}, which is consistent with the age of young massive stars in the Trapezium cluster \citep{1997AJ....113.1733H, 2014AstBu..69...46B}. We have calculated the position of the IF for 0.01, 0.06, 0.1 and 0.3 Myr using equation \ref{eq:rif}. These positions are shown using cyan, magenta, red and black concentric circles respectively, centred at $\theta^1$~Ori~C in Figure \ref{fig:stromgren}. The circles corresponding to 0.01--0.06 Myr radii lie well inside the observed emission boundary, while the 0.1 Myr circle contains most of the radio emission. The 0.3 Myr radius extends beyond the observed boundary. This indicates that ages significantly above $10^5$ yr cannot reproduce the measured size of the ionised region. The comparison therefore agrees best with the age of the ionising stars between 0.06--0.1 Myr. Furthermore, within the calculated classic Str\"{o}mgren radius, the emission is predominantly optically thick free–free radiation, as expected for a dense H~{\sc ii} region. The correspondence between the predicted ionised extent and the observed spectral index morphology provides a self-consistent physical picture of the thermal emission distribution across the nebula.

\subsection{Unambiguous signature of non-thermal emission}
\begin{figure}
    \centering
    \includegraphics[trim={0 4cm 4cm 2cm},clip,width=1.0\linewidth]{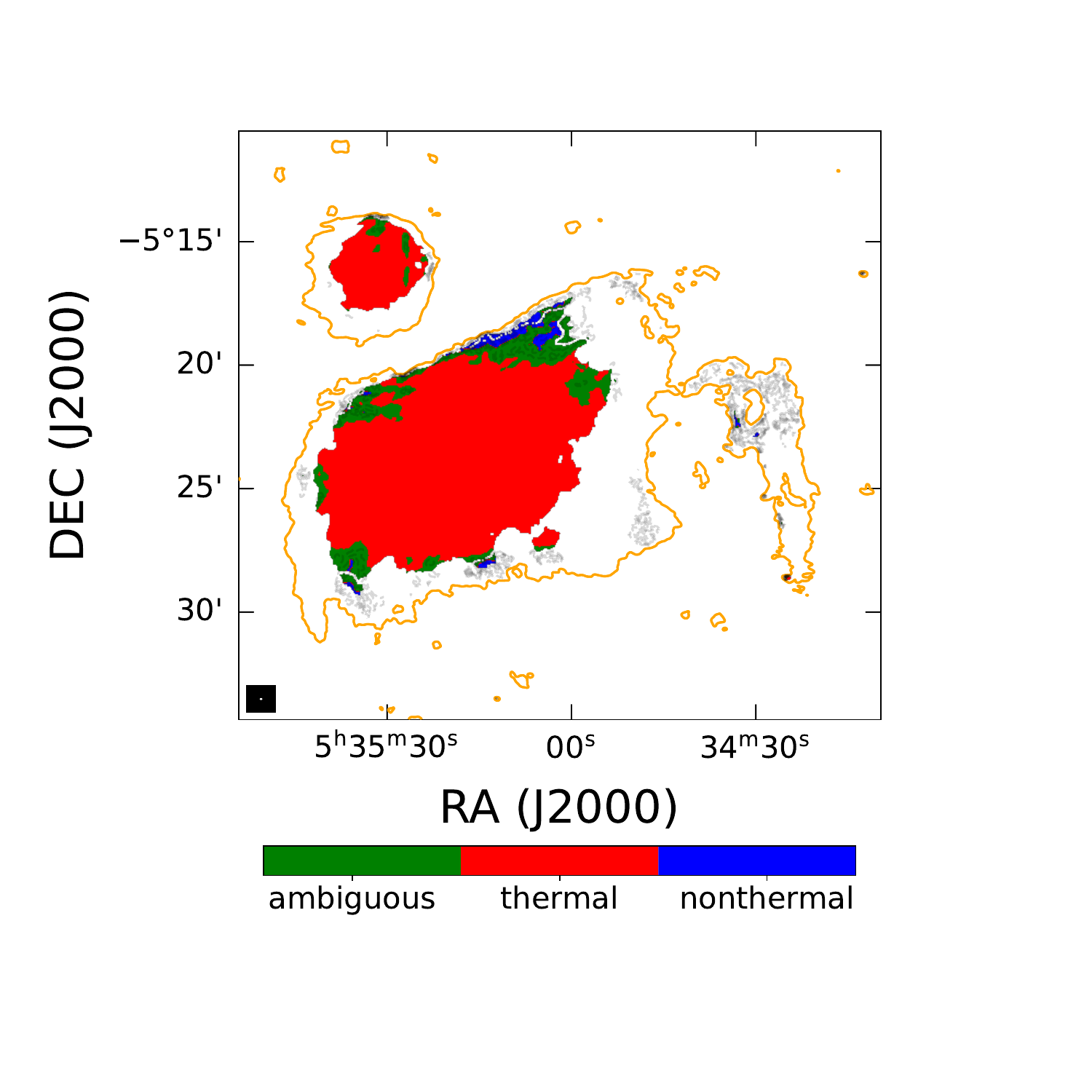}
    \caption{Pixel map classified into high confidence thermal (red), non-thermal (blue), and ambiguous (green) regions. The yellow outlined contour is drawn at S/N=10. Semi-transparent grey regions are contours of low-confidence non-thermal pixels.}
    \label{fig:conf-sp}
\end{figure}    
\par
Theoretically, for radio synchrotron emission, the spectral index lies in the range, $-2 \leq \alpha \leq -0.1$. However, in radio observations, the emission is considered non-thermal if the measured spectral index, $\alpha_{\text{thresh}} \leq -0.4$ \citep{1999ApJ...527..154K}. We considered these requirements, together with constraints from extended source simulation, to determine the high-confidence non-thermal regions in our spectral index maps. Given the asymmetric percentile statistics with bias in low S/N bins (for S/N<68), to declare any emission as non-thermal with certainty ($95\%$), the measured spectral index has to be $\alpha_\text{NT}\leq\alpha_{\text{thresh}}-\alpha_{\text{bias}} +\Delta\alpha _ \text{upper}$ in this S/N range, where $\Delta\alpha_\text{upper}=\alpha_{50} -\alpha_{97.5}$. Similarly, theoretically for the thermal emission $\alpha_\text{thresh}>-0.1$. Hence, in the said range S/N ranges, the value must be $\alpha_\text{T}> \alpha_\text{thresh} -\alpha_\text{bias}+\Delta\alpha_\text{lower}$ to be considered as thermal emission, where $\Delta\alpha_\text{lower}=\alpha_{50} -\alpha_{2.5}$. The measured $\alpha$ values falling between the values derived by the above two criteria in each bin remain ambiguous. Since, for high S/N ranges (for S/N>68), the total deviations themselves are small and fall approximately below 0.2 on either side, we consider a general criterion. We consider pixels to be non-thermal if $\alpha_{NT}<-0.6 (-0.4-0.2)$ and thermal if $\alpha_{T}>0.1 (-0.1+0.2)$. Applying these criteria yields a high-confidence classification of thermal and non-thermal pixels. Figure~\ref{fig:conf-sp} shows a plot based on such criteria. The yellow outline represents the contour of S/N = 10 from the band 4 image. We note that we could not reliably quantify the error in the estimated $\alpha$ for pixels with S/N<25 in our simulation. Thus, even though there exist regions with spectral index characteristic of non-thermal emission in the S/N range of 15 (our cutoff for $\alpha$ map) to 25 adjacent to the high confidence pixel and are part of the same morphological structure (the connection is illustrated using contours of semi-transparent grey colour), we refrain from concluding that the emission from such pixels are unambiguously non-thermal.

\par
Considering the constraints of the simulation and the results from continuum analysis, it is evident that the signature of the non-thermal emission is real and originates from some physical process occurring at the Orion Nebula. Moreover, clustering of the high-confidence pixels together further signifies that they are arising from certain parts of physical morphology and not occurring in a scattered manner like artefacts. Hence, it is highly unlikely that these non-thermal emission are manifestations of artefacts in the radio interferometric data analysis. 

\par
Our high-resolution spectral index measurement broadly agrees with the report of \cite{2001AJ....121..399S}, particularly in the bright central region ($\alpha\sim2$, consistent with optically thick thermal emission) and in significant parts of the outskirts of the central region ($\alpha\sim-0.1$, consistent with optically thin thermal emission). However, we have detected new non-thermal emission from some of the regions in our study. We suggest that, due to the coarse resolution of their map ($80''$), localised regions of non-thermal emission were missed due to the averaging of the emission with the thermal emission.

As discussed earlier, non-thermal emission in H~{\sc ii} regions is neither common nor expected,
since these regions are primarily composed of ionised gas that emits thermal free–free radiation.
However, under certain circumstances, a non-thermal component may arise. In some cases,
non-thermal radio emission has been detected when the shock front of a nearby supernova
remnant interacts with an H II region \citep{2019MNRAS.482.4630V}. The Orion Nebula, however, consists predominantly of a young stellar population and is not known to be associated with
any such supernova activity. Another plausible mechanism involves shocks generated by jets or
outflows from young stellar objects (YSOs). Although non-thermal emission associated with
YSO-driven shocks have been observed in a few cases \citep[e.g., ][]{2018MNRAS.474.3808V}, such detections remain relatively rare. There are also other, less conventional but not entirely dismissible possibilities, such as emission arising from cloud–cloud collisions or from cosmic-ray electrons interacting with local magnetic fields. In the subsequent subsections, we discuss potential
scenarios supported by previous observations that could account for the observed non-thermal emission in Orion, which may serve as hypotheses for future detailed modelling efforts.

\begin{figure}
    \centering
    \includegraphics[width=1.0\linewidth]{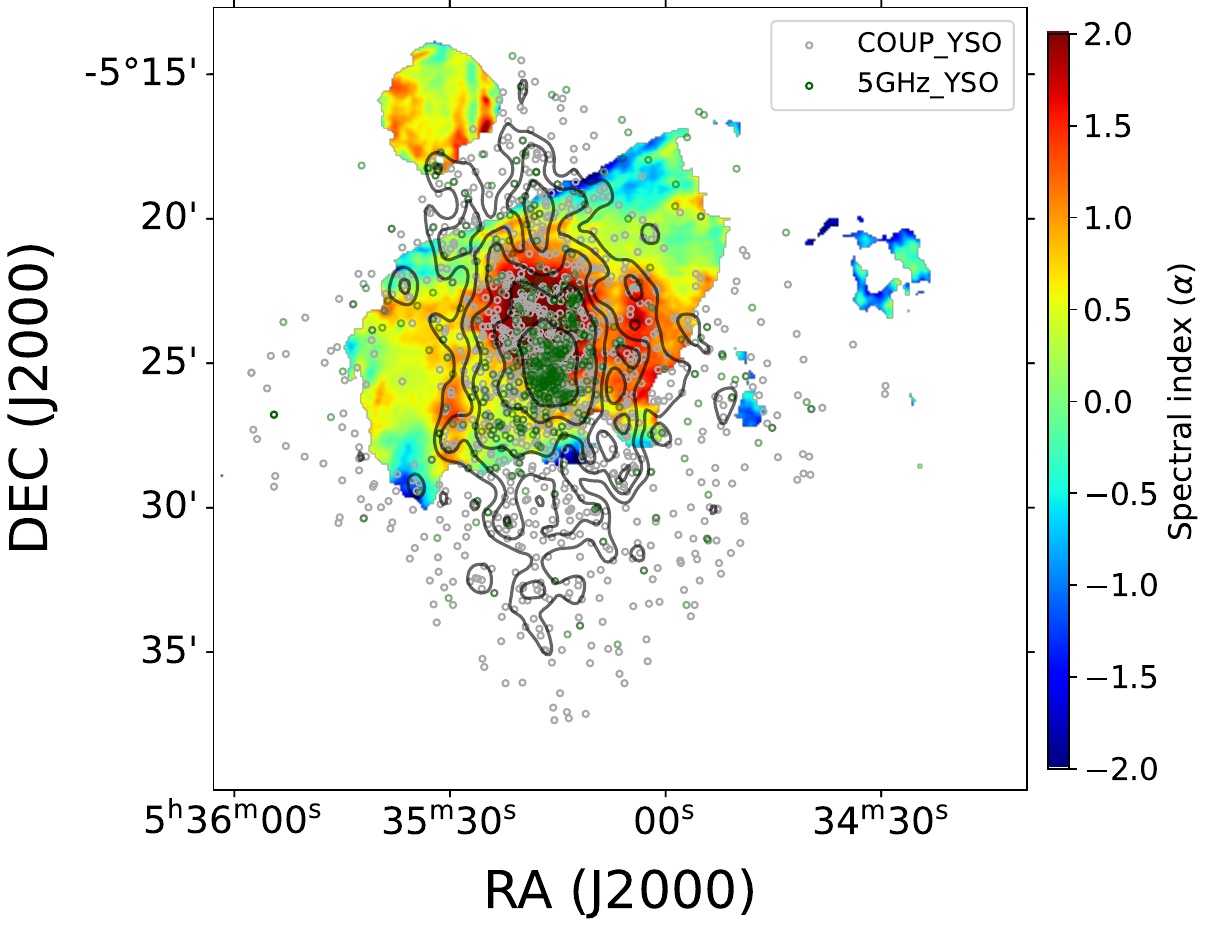}
    \caption{KDE contour plot of the COUP X-ray YSO sources \citep{2005ApJS..160..319G} overlaid on the spectral index map. The coordinates of COUP sources are marked with grey circles, and the 5 GHz radio YSO sources \citep{2021MNRAS.506.3169V} are marked with green circles.}
    \label{fig:YSO-Orion}
\end{figure}

\begin{figure*}
    \centering
    \includegraphics[width=\linewidth]{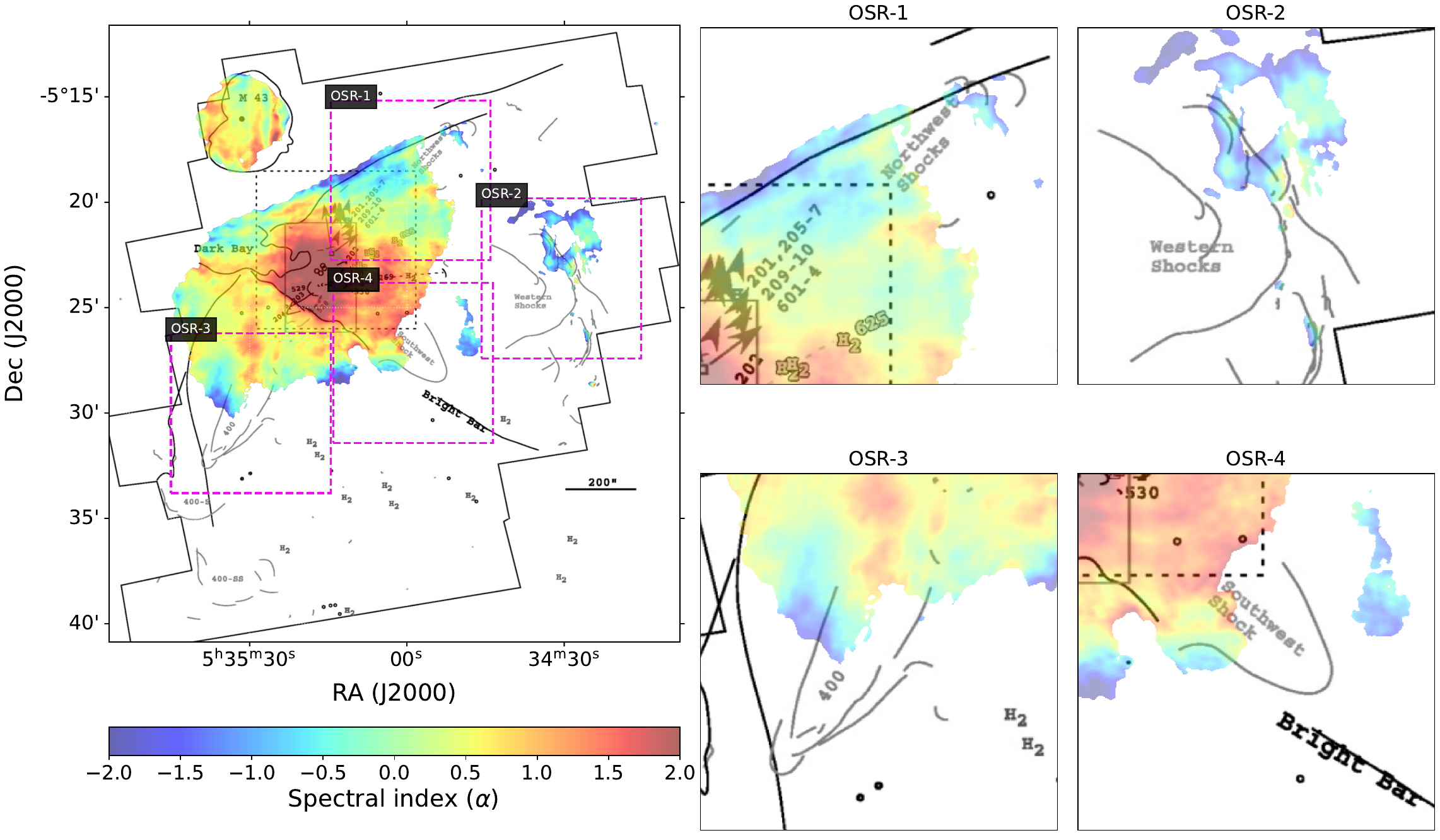}
    \caption{The drawing of shock fronts from optical study \citep{2007AJ....133.2192H} overlaid on the spectral index map (shown in colour) in the \textbf{left}. The irregular semi-rectangular boundary shows the area covered by the HST optical study. The thick black lines are bright boundaries of the Orion nebula known as "Rims" \citep{2010AJ....140..985O}. Lighter grey lines trace the shock features detected in the optical study. In the \textbf{right} sub-panels, four prominent shock fronts overlapping with the non-thermal radio emission are magnified.}
    \label{fig:henry_hst}
\end{figure*}
\begin{figure*}
    \centering
    \includegraphics[width=16cm]{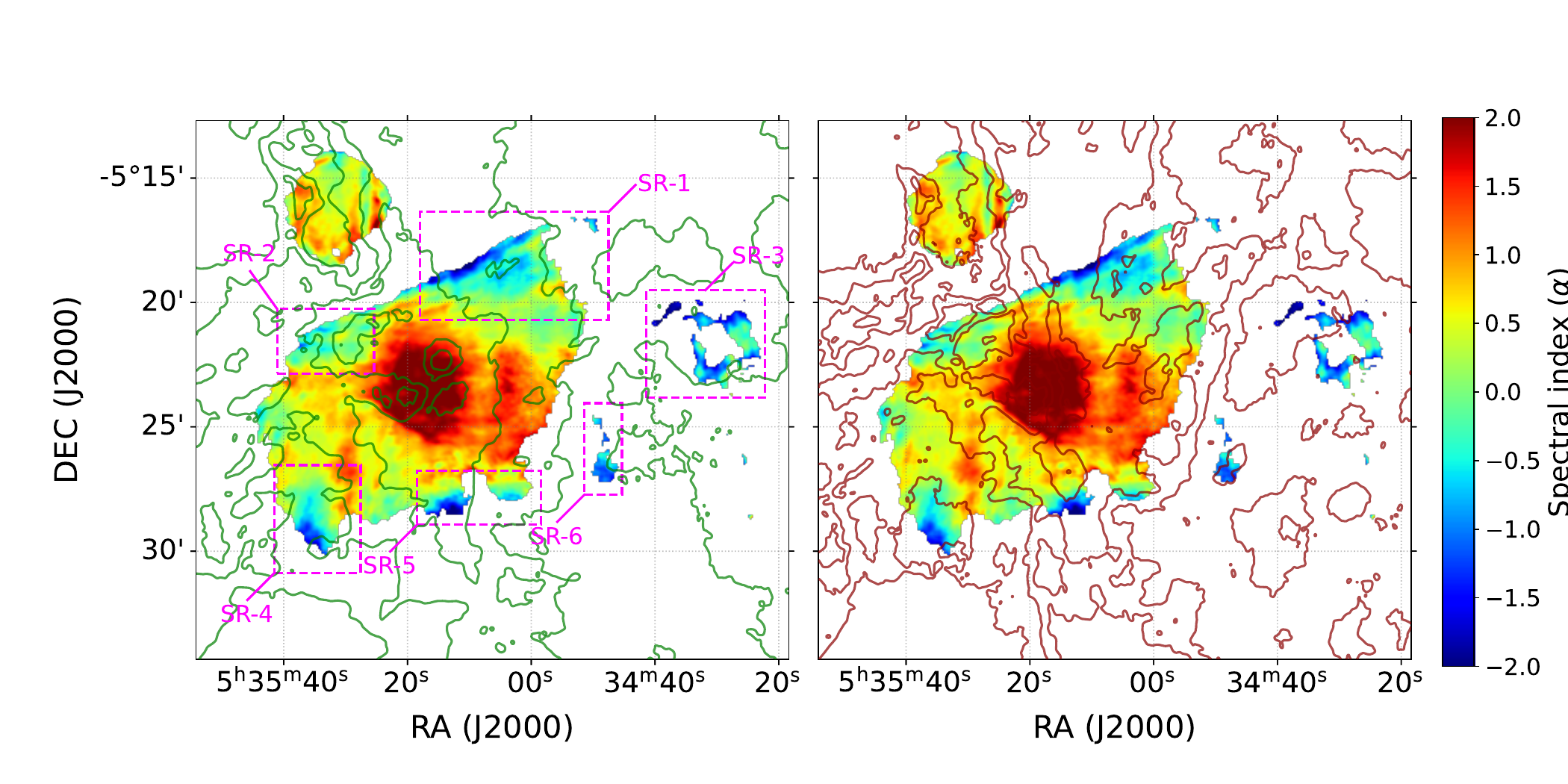}
    \caption{$^{12}$CO \textbf{(left)} and $^{13}$CO \textbf{(right)} zeroth-moment contours from the CARMA NRO Orion survey \citep{2018ApJS..236...25K} are overlaid on the spectral index map. The contours are at [7, 14, 21, 42, 70] $\%$
    of the peak brightness of the respective image zeroth moment map. }
    \label{fig:Orion-CO}
\end{figure*}

\subsection{Young star formation activity \& stellar outflows}
The process of star formation results in the creation of outflows, jets, shocks and stellar winds, all of which inject energy, leading to turbulence in the surrounding ISM \citep{2001ARA&A..39...99O}. This can either disperse the gas in the medium or induce further star formation, the latter potentially creating a self-sustaining feedback loop. While the presence of stellar nurseries is well established through NIR observations, detecting radio non-thermal emission provides complementary insight into the energetic feedback from jets and shocks, which play a key role in shaping the ISM during active star formation.
\par
Figure~\ref{fig:YSO-Orion} shows the KDE plot (Scott's rule, adjustment factor=0.35) of YSOs (detected in X-ray band) spatial distribution overlaid on the spectral index map. The X-ray catalogue is taken from the Chandra Orion Ultradeep Project \citep[COUP;][]{2005ApJS..160..319G}; one of the most reliable YSO catalogues available. We have shown the positions of the COUP X-ray sources using grey circles. We have also shown the positions of the YSOs taken from the 5~GHz VLA study of the Orion region by \citet{2021MNRAS.506.3169V}; most of the 5~GHz YSO objects are the counterparts of the COUP X-ray sources. In line with expectations, the central bright region near the Orion bar hosts most of the YSOs. Away from this region, the density of the YSOs gradually decreases. However, some local overdensity of these objects can be seen in the SR-1, SR-2, SR-4, and SR-5, as well as in the vicinity of the SR-6 region of Figure~\ref{fig:spectra-index-orion}. However, these overdensities are too weak compared to the density of YSO in the central region, and hence they can not be conclusively associated with the synchrotron emission.

In Figure~\ref{fig:henry_hst}, we show an overlay of a drawing of numerous shock fronts associated with the HH objects \citep{2001ARA&A..39..403R} on the spectral index map, from the optical study by \citet{2007AJ....133.2192H} using the Advanced Camera for Surveys (ACS) of the Hubble Space Telescope (HST). Their work shows that these shocks originate from the Optical Outflow Source (OOS) located at $\sim1'$ southwest of the $\theta^1$ Ori \textit{C}. In the left panel of Figure~\ref{fig:henry_hst}, we show the overall overlay of the optical drawing with spectral index map and the four sub-panels (OSR-1, OSR-2, OSR-3, OSR-4) at the right magnify the optically detected shock features. OSR-2 shows non-thermal regions within the shell-like feature from SR-3 (about $\sim5'$ across; see Figure \ref{fig:spectra-index-orion}), and some pixels a few arcminutes south of it are seen to spatially overlap with the Western shocks. In OSR-3, the non-thermal feature of $\sim2'$ in length (along the larger dimension) close to the southeast direction is also seen to overlap with the HH-400 object, which is known to have a velocity in the plane of sky \citep{2001AJ....122.1508B}. OSR-1 shows a triangular-shaped (height $\sim2'$) non-thermal region of SR-1, which has a partial coincidence with the northwestern shock features. In OSR-4, a small region of non-thermal emission ($\sim30''$) of SR-5 also partially coincides with the right side of the southwestern shock. However, we found no shock front associated with the non-thermal emission shown in SR-2 and SR-6.
\par
Recent studies of [C~{\sc ii}] emission \citep{2021A&A...652A..77H} have identified "protrusion" \citep{2022A&A...660A.109K} and "dents" \citep{2022A&A...663A.117K} on the Orion's veil. These studies have established that such features result from the interaction between the stellar outflows and the large-scale expanding bubble driven by the stellar winds from Trapezium stars (most likely by $\theta^{1}$~Ori C). However, these features remain undetected in CO studies, suggesting a low molecular gas fraction or consisting of CO-dark H$_2$. The "protrusion" consists of two half-bubbles ($\sim4'$ and $\sim8'$ across) expanding towards us with an expansion velocity slower than the Veil, as evident from the PV diagram (Figure 3) shown in \citet{2022A&A...660A.109K}. They have suggested that these features are formed by fossil bipolar outflows of trapezium star(s). The "protrusion" extends between our SR-1 and SR-3 regions. Furthermore, \citet{2022A&A...663A.117K} suggest six identified "dents" are the shock-accelerated gases created on the veil shell by the jets and outflows from the massive stars. Interestingly, some of these features are also located near or spatially coincide with the non-thermal regions. The location of the dents on the spectral index map is shown in Appendix \ref{fig:dent}. D1 is close to SR-5, D2 and D4 are close to the SR-6 region, and D3 is partially overlapping with SR-3. However, our observation has not recovered any radio emission from the positions of D5 and D6 dents. In addition, the structure of the veil shell is complicated, and newer components are still being characterised in recent studies \citep{2025AJ....169..252O,2020A&A...639A...2P, pabst2019disruption}. We will explore the connection of the veil shell with non-thermal emission in further detail in Rashid et al. (in preparation).      
\par
The spatial overlap between optically identified shocked regions and non-thermal radio emission provides compelling evidence for an association between the non-thermal emission and stellar outflows, particularly Herbig-Haro (HH) objects. The presence of distinct velocity structures in [C~{\sc ii}] emission near these regions further strengthens this interpretation. However, the velocities of the gas components are too low to account for the synchrotron emission observed in our study. Moreover, the absence of shock features in SR-2 and SR-6, along with very small-scale optical shock fronts in SR-1, suggests that shocks associated with stellar outflows may not be the only process responsible for non-thermal emission in the Orion nebular region. Additionally, the shock features from the optical studies extend far beyond the morphology of the radio continuum emission of our study, and our observation is not deep enough to recover all the faint non-thermal emission surrounding all the shock features. The limited sensitivity complicates the efforts to characterise the morphology of the shocked region in the radio regime. Since the flux densities of non-thermal emission are expected to be higher at lower frequencies, future observations at even lower frequency bands could provide additional insights, though the potential degradation in resolution remains a concern.

\subsection{Collision of molecular clouds}
\begin{figure*}
    \centering
    \includegraphics[width=0.45\linewidth]{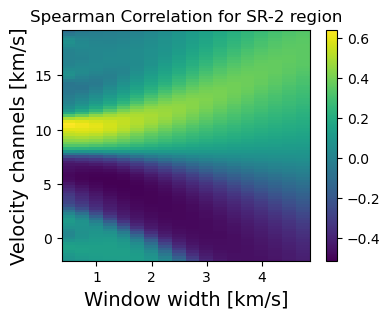}
    \includegraphics[width=0.45\linewidth]{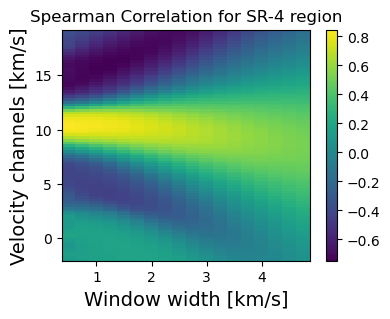}
    \caption{Spearman rank correlation maps between spectral index and Gaussian weighted integrated intensity over different velocity windows centred at particular velocity channels. \textbf{Left} panel is for SR-2 regions and \textbf{Right} panel is for the SR-4 region. The X-axis is the width of the velocity window (in km/s) of the Gaussian filter used to calculate zeroth moments. The Y-axis is the central velocity channel. Colour represents the values of the Spearman coefficient.}%
    \label{fig: corr-plt}
\end{figure*}
The supersonic collision between two interstellar molecular clouds in the star-forming region may create a violent environment. This can cause shocks in the ISM. Recent studies have explored both observationally and theoretically the impact of such collisions on high-mass star formation \citep{2014ApJ...792...63T, 2015MNRAS.450...10H, 2017ApJ...837...44D, 2017ApJ...835..142T, 2018ApJ...859..166F}. \cite{2013ApJ...774L..31I} showed in their magneto-hydrodynamic (MHD) simulation that an enhanced magnetic field and turbulence can result in a larger effective Jeans mass in the molecular clouds, which can serve as a seed for the formation of massive stars. The shock-accelerated electrons in such magnetised clouds may manifest non-thermal emission.

\par The rotationally excited CO lines trace molecular hydrogen. Figure~\ref{fig:Orion-CO} shows the overlay of $^{12}$CO (left panel) and $^{13}$CO (right panel) velocity integrated intensity (zeroth-moment in the range of 2.5–15km~s$^{-1}$) contours, taken from CARMA NRO Orion survey \citep{2018ApJS..236...25K}, on the spectral index map. These maps show local peaks of cold molecular gases spatially coinciding with some non-thermal emission regions. However, the spatial overlap of non-thermal regions in the spectral index map with the integrated emission (across all the velocity channels) or zeroth-moment map does not guarantee that velocity-separated gas components (signature of collision) are correlated to the non-thermal regions, as all the components are summed.
\par
We carried out a correlation analysis between the spectral index map and Gaussian weighted zeroth moment maps to identify correlated velocity-coherent gas structures with the non-thermal regions. We first cropped the SR subregions (as in Figure \ref{fig:spectra-index-orion} and \ref{fig:Orion-CO}) from both the $^{12}$CO spectral cube of \cite{2018ApJS..236...25K} and the spectral index map, and convolved them to a common angular resolution. Next, we applied Gaussian filters with velocity windows ranging from 2 to 20 velocity channels ($\sim$ 0.5 to 5 km~s$^{-1}$) in width with one channel increments ($\sim0.25$ km~s$^{-1}$) to the cube to bring out velocity-coherent structures. For each Gaussian-weighted cube, we compute the zeroth-moment map, which enhances structures coherent in velocity space. These maps represent the spatial distribution of emission corresponding to a particular velocity window. We then compute the Spearman's rank correlation coefficients \citep{ca468a70-0be4-389a-b0b9-5dd1ff52b33f} between the zeroth-moment map and the spectral index map. Examples of cropped SR-2 region from the spectral index map and the moment zero map are shown in Appendix \ref{fig:crop24}.

\par We produced the Spearman coefficient's map as a function of channel and width of the Gaussian velocity window (WD). We observe distinct patterns in these maps across different sub-regions, revealing varying degrees of spatial association between the integrated $^{12}$CO emission and the radio spectral index. The peak absolute values of the Spearman coefficient for each subregion lie within the range of $\sim0.42$–$0.85$, indicating moderate to strong correspondence across the regions. Figure \ref{fig: corr-plt} shows Spearman coefficient maps of two maximally correlated regions (SR-2 and SR-4). These correlations are calculated between the spectral index map and the Gaussian-weighted zeroth moment map for different WDs. The colour represents the values of Spearman's coefficients. The maps show multiple correlated and anti-correlated branches centred at different velocity channels representing gas components integrated over different WD. The correlation map for the SR-2 region (left) shows two gas components (centred at $\sim6.81$ km~s$^{-1}$ and $\sim10.56$ km~s$^{-1}$) marginally separated in velocity space have moderate to strong correlation (absolute Spearman coefficient $\sim0.5-0.62$) with the non-thermal region for WD of 2–10 velocity channels. The gas component at 6.81 km~s$^{-1}$ shows a negative correlation with spectral index values $\alpha$, signifying increasing intensity of $^{12}$CO emission with increasing negative value of $\alpha$, that is, with non-thermal emission. We have shown the channel map around this component in Appendix \ref{fig:co_chan} as an example case to visualise the emission. The gas component at 10.56 km~s$^{-1}$ shows a positive correlation, i.e. increasing $^{12}$CO intensity with increasing values of $\alpha$. The map of the SR-4 region shows three gas components (centered at $\sim5.56$ km~s$^{-1}$, $\sim10.56$ km~s$^{-1}$, and $\sim15.57$ km~s$^{-1}$) for a similar WD range, the latter two of which have strong correlation (absolute Spearman coefficient $\sim0.6$–$0.82$). We observe a similar trend of correlation and anti-correlation between $\alpha$ and the $^{12}$CO intensity of different gas components. The correlation plots for the other SR regions are shown in Appendix \ref{fig:corr_all}. While these regions also exhibit multiple gas components, the correlation tends to be moderate. Note that the anti-correlation of $^{12}$CO intensity with the negative spectral index $\alpha$ signifies the correlation with the non-thermal emission as $S\sim\nu^\alpha$. 
\par
The presence of molecular gas components that are marginally separated in velocity space, yet appear co-spatial in projection, is a commonly used observational signature of cloud-cloud collisions. Such configurations suggest that the clouds are interacting along the line of sight, and their physical overlap in the sky plane reinforces the possibility of a past or ongoing collision. \cite{2018ApJ...859..166F} hypothesised that the two velocity components of the Orion A molecular cloud collided to trigger the formation of massive stars in M42 and M43. They reported that the two complementary features have a relative velocity of $\sim6$~km~s$^{-1}$, and the separation between the clouds is estimated to be $\sim0.5$~pc. Furthermore, they discussed that a depression in intensity is expected at cloud collision sites in the optical/infrared bands. The HST ACS images show many obscured regions in the Orion region \citep{2013ApJS..207...10R}. For example, the Dark Bay and certain regions of the Western Shocks are highly obscured regions. They suggested a possibility of a cloud collision; however, they also outlined several caveats, suggesting that the hypothesis may not be conclusive.    
\par 
The redshifted cloud component (12.9-14.9 km~s$^{-1}$) shown in \citet{2018ApJ...859..166F} has clumps that are spatially coincident with the non-thermal emission in the SR-1 and SR-2 regions of this study. The blue-shifted component at 8.8 km~s$^{-1}$ has co-spatial clumpy features with the non-thermal emission in the SR-3, SR-4, SR-5, and SR-6 regions. However, we do not see strong signatures of collision or turbulence in all of the non-thermal regions from our correlation study. This makes the hypothesis that the non-thermal emission arises from collisions between gas components less certain. However, determining the signature of the collision in velocity space is subject to inclination angle \citep[see Figure 2 \& 7 of ][]{2018ApJ...859..166F}. Therefore, the collision scenario cannot be entirely ruled out.
\par

Turbulence in the gas components can be measured by analysing line widths. In this study, we did not directly correlate the line width with the spectral index values. In the near future, we will carry out a study correlating the turbulence in the gas with the non-thermal region in the companion paper, Rashid et al. (in prep), which may help firmly establish (or rule out) the cloud collision scenario. Here, we have shown average $^{12}$CO spectra from different parts of the Orion nebula in Appendix \ref{fig:co-spectra}. The width of the spectrum is the widest at the centre of the cloud. However, the spectra away from the central region, where non-thermal emission is not detected, are narrower and seem to have a single component. In contrast, the spectra from the SR-2 region are broader and multi-component, and hence could be more turbulent.       
\par
However, the velocity components involved in the above interpretation are too low to accelerate electrons locally. We suggest that the gas density is enhanced due to the sweeping up of material by the correlated velocity components, and is highly magnetised as a result. These magnetised regions can trap the existing high-energy electrons from the Galactic cosmic ray population, resulting in synchrotron radiation. In an alternative scenario, the non-thermal emission could arise from the expansion of the wind-driven bubble into the dense molecular medium. The interaction of the veil layers at the PDR interface may result in an enhanced gas density, which in turn leads to an enhanced magnetic field. The anticorrelation of $\alpha$ with CO emission, i.e. enhancement of CO intensity with non-thermal emission, possibly signifies this phenomenon. Direct measurements of magnetic fields in molecular gas are notoriously difficult and, in many cases, practically unfeasible. Magnetic fields can be probed using H~{\sc i} Zeeman splitting and dust polarisation measurements; however, these methods do not yield precise estimates of the magnetic field strength within molecular clouds. Since magnetic field strength is expected to scale with gas density \citep[e.g., B$\propto \mathrm{n}^\kappa$, where $\kappa\sim0.5$; ][]{2012ARA&A..50...29C}, the magnetic field in dense molecular gas should be stronger than that inferred from H~{\sc i} Zeeman measurements, which typically trace lower-density atomic regions. \cite{2016ApJ...825....2T} have studied the line-of-sight (LOS) magnetic field from Zeeman effect using H~{\sc i} and OH data. Their study has mapped the LOS magnetic field from two principal H~{\sc i} components at 3 (B) and 6 (A) km~s$^{-1}$. Both components show enhancement toward the northeast region of their map \citep[see Figure 5 \& 6 of ][]{2016ApJ...825....2T}. Due to masking pixels $<3\sigma$ uncertainty in the LOS magnetic field, they could not map the entire EON region. However, they cover the SR-2 region in the component A map, and their map shows an enhanced magnetic field in the overlapping region than in the central region. We note that we could not find a well-suited polarisation study to compare with our results. We further discuss the wind-driven feedback in the context of studies of high-energy emission in the next section.

\subsection{Cosmic rays acceleration by stellar winds of massive stars}
\begin{figure}
    \centering
    \includegraphics[width=1\linewidth]{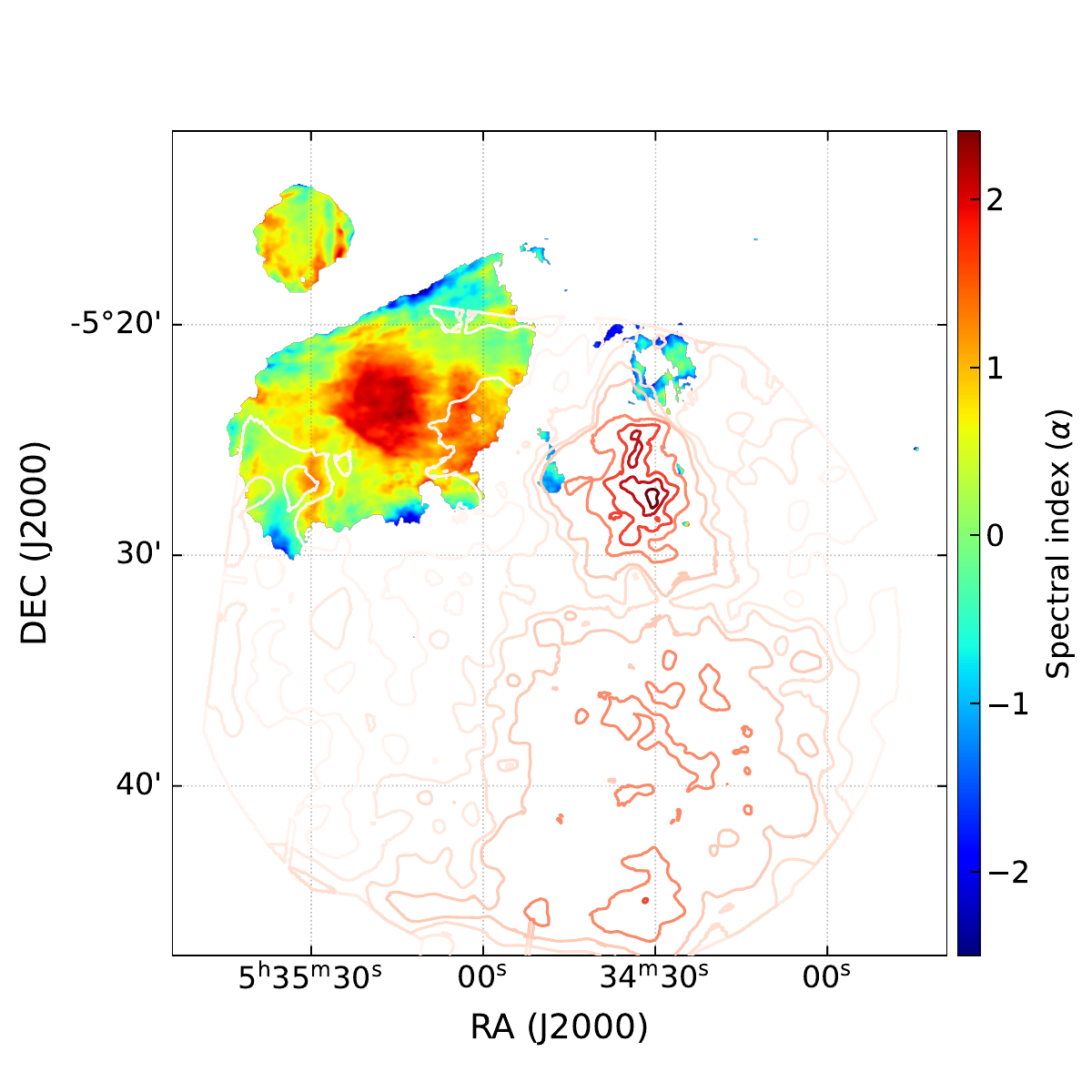}
    \caption{Contours from diffuse X-ray emission from XMM Newton telescope \citep[from ][]{2008Sci...319..309G} overlaid on the spectral index map.}
    \label{fig:xray}
\end{figure}
The Orion Nebula is the nursery for massive stars. It hosts a large number of O/B type stars \citep[see Figure B1 of ][ for a map of massive stars in the Orion region]{2022A&A...663A.117K}. The energetic winds from these stars can result in shock compression of their surroundings. These shocked layers can accelerate cosmic rays, and their signatures can be seen in high-energy bands like X-rays and Gamma rays or in radio emission. Recent studies have discussed the observational and modelling aspects of this scenario \citep{2022MNRAS.510.5579B,2016A&A...591A..71M,2024NatAs...8..530P}. Here, we discuss results from X-ray and gamma-ray studies of the Orion Nebula and the possible connection with the non-thermal radio emission.
\citet{2008Sci...319..309G} reported diffuse X-ray features filling in the infrared structures formed by shocks in the western part of the EON. The diffuse X-ray emission was attributed to hot plasma (1.7–2.1 million K) produced by shock-heating from high-velocity stellar winds of Trapezium stars, particularly $\theta^1$ Ori C. They showed that the kinetic energy rate of these winds ($\sim 7 \times 10^{35}$ erg/s) far exceeds the observed X-ray luminosity ($\sim 5.5 \times 10^{31}$ erg/s), suggesting winds as the dominant heating mechanism. Furthermore, they suggested, in near-pressure equilibrium with cooler H~{\sc ii} gas, the plasma escapes the cavity as an "X-ray champagne flow," feeding much larger interstellar structures like the Eridanus superbubble. \citet{2008Sci...319..309G} also argued against alternative sources, such as unresolved stars or supernova remnants, as the origin of the diffuse X-ray emission—on the basis of the soft spectral nature of the emission, the absence of a dense stellar population in infrared observations, and the lack of radio structures typically associated with supernova remnants—further reinforcing the role of stellar winds in driving the diffuse X-ray structures. Such environments are potential sites for cosmic ray accelerations suggested by recent studies \citep[e.g.][]{2024NatAs...8..530P}. In addition, the leptonic cosmic rays may contribute to the non-thermal radio emission in star-forming regions and can be detected in low radio frequencies \citep{2018A&A...620L...4P}. \par
Figure~\ref{fig:xray} shows an overlay of the X-ray emission \citep[obtained by private communication from ][]{2008Sci...319..309G} on the spectral index map. The X-ray emission in their observation is strongest toward the southwest of M42 and does not fully coincide with the radio non-thermal emission, appearing only in the vicinity of the bubble (e.g., SR-3 and SR-6 regions). However, the X-ray observation does not cover the SR-1 and SR-2 regions. A deeper and comprehensive X-ray map of the region could offer further insights. Interestingly, Chandra observations have not detected any diffuse X-ray emission in the brighter central parts of M42 \citep{townsley200310}. We also note that radio emission remains faint toward the southern and southwestern parts of the EON, and it is unclear whether additional non-thermal features exist that correlate with the structure of the bubble in a manner similar to the SR-3 region. One possible explanation is the tenuous nature of the plasma ($\sim 1\text{ cm}^{-3}$) inside the bubble, with non-thermal radio emission becoming stronger only at the periphery, where there is an accumulation of swept-up material \citep{pabst2019disruption}. While the connection between the radio non-thermal emission and the hot plasma on larger scales remains uncertain, such a scenario cannot be ruled out. \par

Gamma-ray emission is smoking-gun evidence of the hadronic cosmic-ray signature. However, gamma-ray studies are limited by their relatively poor angular resolution, with the Fermi Large Area Telescope (LAT) having a resolution of 0.1 degrees at energies beyond 1 GeV. It is, hence, challenging to localise any emission features to angular scales of a few arcminutes. However, it is to be noted that both Fermi and AGILE have detected Gamma-ray excess from the Orion A molecular cloud at large scales \citep{2012ApJ...756....4A, 2018A&A...615A..82M}.      

\section{Conclusions}
\label{sec: conclusion}
In this paper, we have utilised the low-frequency (<1GHz) uGMRT data to conduct the radio continuum study of the Orion Nebular region and produce a high-quality continuum and spectral index map. We have tested the reliability of spectral index values by simulating the visibility data in the uGMRT configuration. Based on our study, we draw the following conclusions:
\begin{enumerate}
        \item We have obtained deep continuum images centred at 400 MHz (band 3) and 685 MHz (band 4) from uGMRT data of the Orion region. The off-source rms noises of the images are 400 and 200 $\mu$Jy-beam$^{-1}$ respectively. These are the deepest images available in these frequency bands, showing the morphological structures of the Orion Nebular region at sub-GHz frequencies.
        \item From the test of the reliability of the two-band spectral index map for extended source at different S/N using simulated uGMRT data, we have constrained the spread in spectral index measurement. We find that in low S/N ranges the spectral index has large uncertainties but for the S/N range greater than 68 the spread roughly fall below $\pm 0.2$.
        \item Radio spectral index analysis of the Orion region reveals unambiguous signatures of non-thermal emission from the peripheral regions of the Orion Nebula-- which is uncommon in the H~{\sc ii} regions-- only a handful of such observations have been reported for other H~{\sc ii} regions.
        \item We have compared the location of non-thermal emission with sources at other wavelengths, ranging from submillimetre to gamma-ray, to explain the possible origin of the non-thermal emission. We have explored three plausible scenarios responsible for the radio non-thermal emission-- the shock fronts of jets/outflows from YSOs created during star formation processes, cloud-cloud collision resulting in shock compression of gas, and mechanical feedback from the winds of massive stars. We have compiled and examined existing observational evidence, discussing the strengths and limitations of each scenario.  
    \end{enumerate}
\par
The upcoming SKA facility will revolutionise the field of radio astronomy with unprecedented sensitivity and resolution, along with an extensive range of frequency coverage. To prepare for the SKA era, leveraging the existing telescopes to determine the limits and opportunities is essential. In our subsequent study, we plan to do a deeper radio observation with uGMRT and MeerKAT and map the entire EON in high resolution to unravel more such non-thermal components by spectral index analysis. A sensitive high-resolution map from multiple frequency bands will allow us to retrieve any weak large-scale non-thermal component and help us to ascertain the origin of such emission when combined with multi-wavelength probes. Moreover, we plan to extend a similar study to other star-forming regions of the Galactic plane. Radio observations for the Metrewave Galactic Plane Survey with uGMRT \citep[MeGaPluG;][]{2023A&A...678A..72D} are ongoing in phases, and are expected to facilitate the identification of additional non-thermal components across the Galactic plane.   


\section*{Acknowledgements}
This work is dedicated to the memory of Prof. Karl M. Menten, who passed away on December 30, 2024. Discussions with and input from him during the initial phase of the project have been extremely valuable. We thank the anonymous reviewer for their careful review, which has significantly improved the manuscript. MR thanks Prof. Bhaswati Mookerjea and Prof. Jan Forbrich for their valuable comments and suggestions during his thesis review, which also helped improve this manuscript. NR acknowledges support from the United States-India Educational Foundation through the Fulbright Program. We are also grateful to Prof. Manuel G\"{u}edel for sharing the X-ray image from their study. This research has made use of NASA’s Astrophysics Data System. We acknowledge Max-Planck-Gesellschaft (MPG) for funding through the Max Planck Partner Group (2019- 2024) to carry out this research. 

\section*{Data Availability}

The data underlying this article will be shared on reasonable request to the corresponding author. The GMRT data underlying this article can
be obtained from the GMRT Online Archive (\url{https://naps.ncra.tifr
.res.in/goa/data/search}) using the proposal id: 31\_106 \& 36\_014. 
 

\bibliographystyle{mnras}
\bibliography{orion}



\appendix

\section{In-band spectral index map}
\label{app:a}
\begin{figure}
    \centering
    \includegraphics[width=\linewidth]{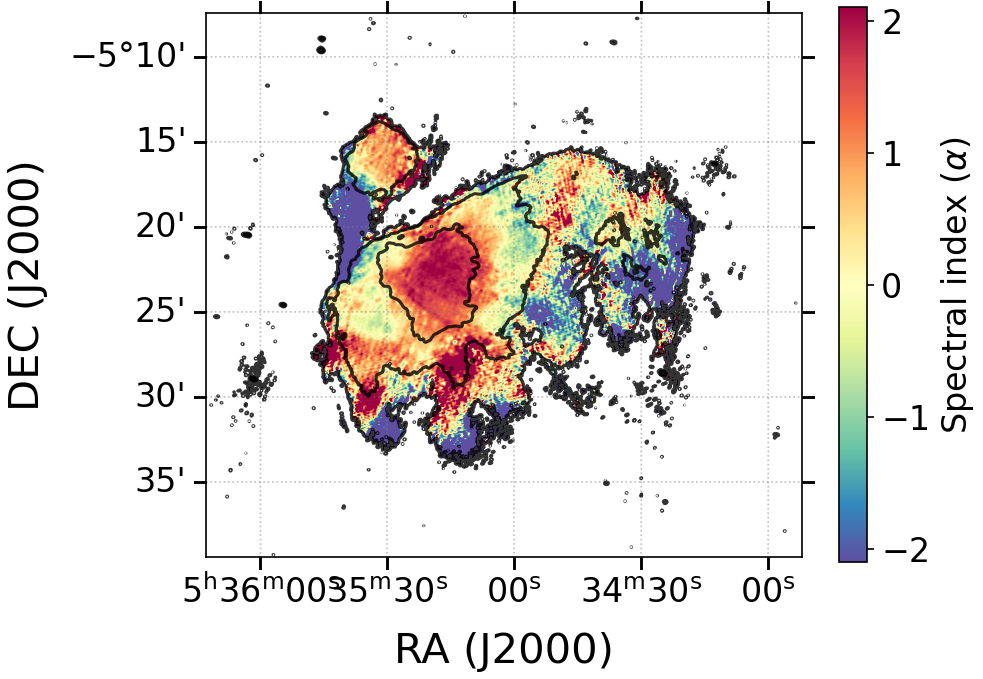}
    \caption{In-band spectral index map from band 3 generated using MTMFS algorithm. The contours are at [1, 5, 30] $\times 3\sigma$.}
    \label{fig:inb-si}
\end{figure}

Figure \ref{fig:inb-si} shows the MT-MFS in-band spectral index map produced from band 3 data of uGMRT. We have used two Tailor terms (nterm=2) to produce this map. The contours are drawn at [1, 5, 30] $\times 3\sigma$. The central region with high S/N shows spectral index values close to +2, consistent with optically thick thermal emission. Away from the central region, the S/N gradually drops to zero. However, some regions show negative spectral index values. Some of these values belong to reasonably high S/N (region enclosed between 15 and 90 times $\sigma$). Moreover, some of the low S/N region also shows extreme negative values, which are unphysical and affected by spurious retrieval of the spectral index.

\section{$\alpha$-S/N distribution of the EON}
\label{app:alpha-SN}
\begin{figure}
    \centering
    \includegraphics[width=0.99\linewidth]{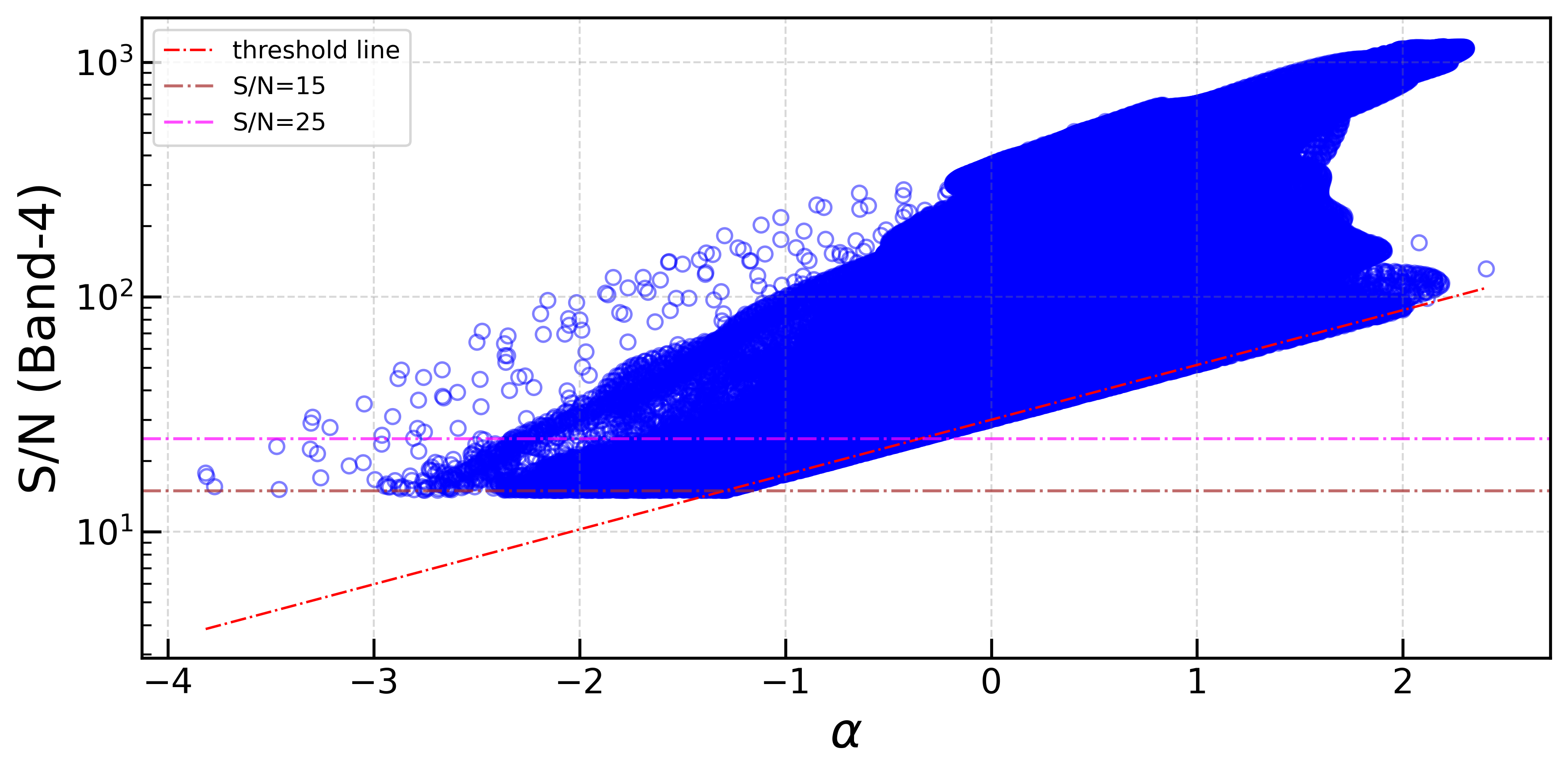}
    \caption{Scatter plot of S/N (band-4) vs $\alpha$ from spectral index map of the EON. The threshold boundaries of $\alpha$ estimation due to the S/N cutoff are shown in red and brown dashed lines. S/N=25 is shown using a dashed magenta line.}
    \label{fig:al-sn-orion}
\end{figure}
Figure~\ref{fig:al-sn-orion} presents the distribution of $\alpha$ as a function of the S/N measured in uGMRT band-4 image of the Orion Nebula. The spectral index is computed only for pixels above a fixed threshold (S/N=15) in both bands. In this parameter space, the detection requirement in Band-3 translates into a curved lower envelope in the $\alpha$-S/N plane, given by
\begin{equation}
    \mathrm{S/N}_{b4} >
N \frac{\sigma_{b3}}{\sigma_{b4}}
\left(\frac{\nu_{b4}}{\nu_{b3}}\right)^{\alpha} .
\end{equation}
In logarithmic scaling of S/N, this equation manifests as a truncation of value below a slanted line, as shown in the plot by the red line and below the horizontal line of S/N=15 shown using a brown dashed line.  
\section{Dents in the veil}
\label{app:veil}
\begin{figure}
    \centering
    \includegraphics[width=1\linewidth]{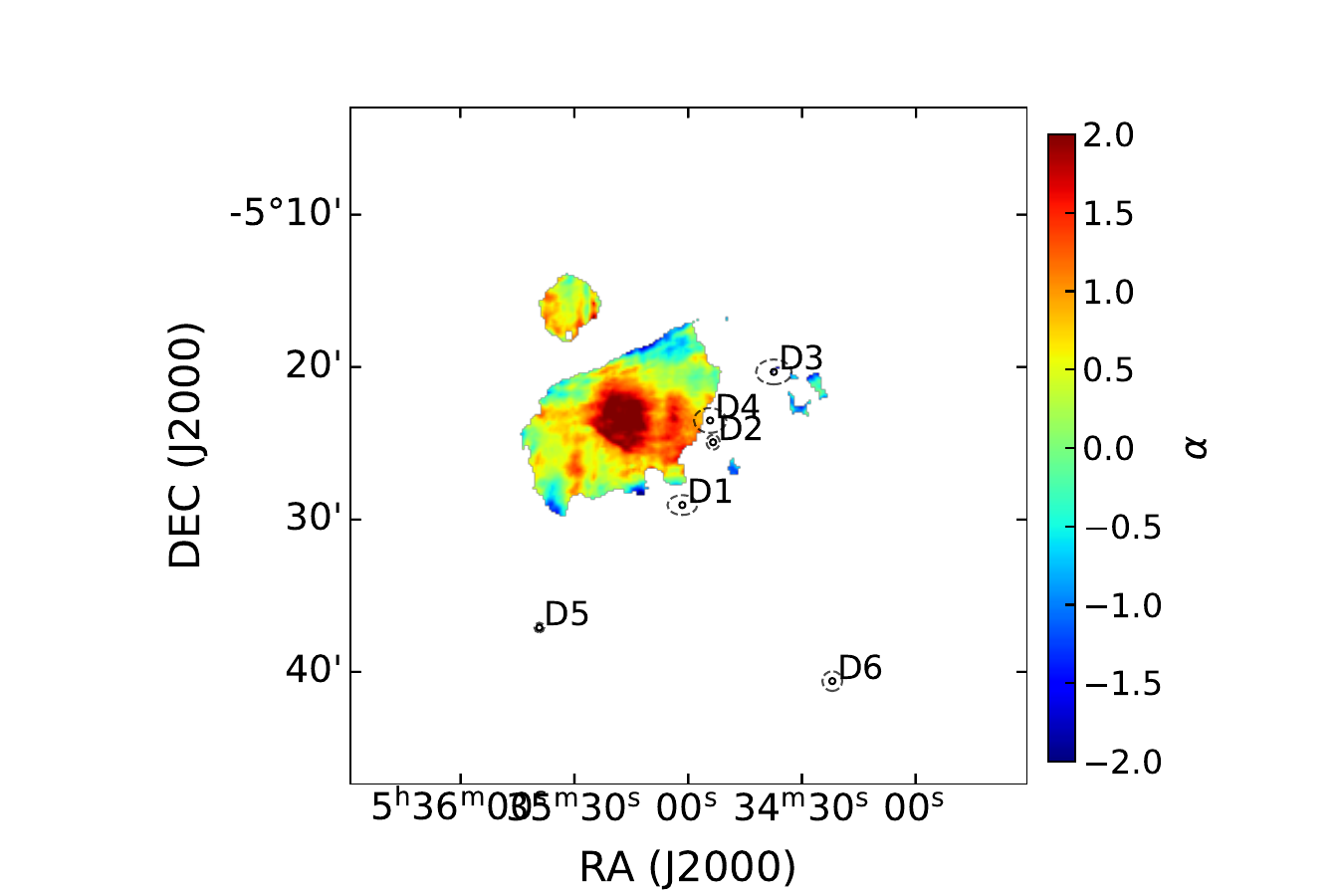}
    \caption{Location of the dents in the Orion region.}
    \label{fig:dent}
\end{figure}

Figure \ref{fig:dent} shows an overlay of the location of the "dents" detected by \cite{2022A&A...663A.117K} on the spectral index map. The dashed ellipses show the approximate size of these features as reported in their work. 
\section{Correlation plots}
\label{app:b}
\begin{figure*}
    \centering
    \includegraphics[width=0.8\linewidth]{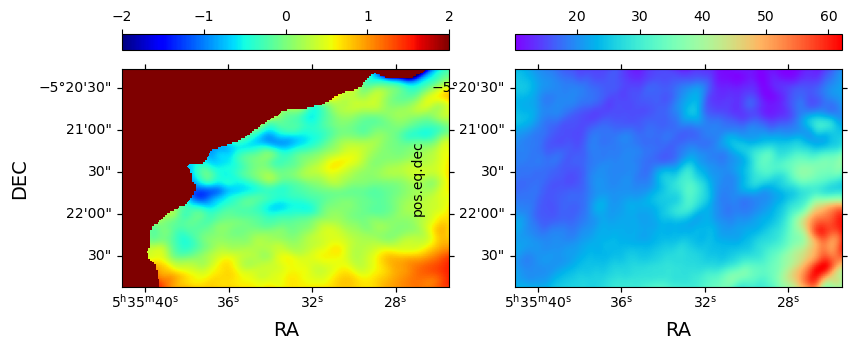}
    \caption{Cropped spectral index map (left) and zeroth moment for entire velocity range from CO emission map (right) for SR-2. }
    \label{fig:crop24}
\end{figure*}

Figure \ref{fig:crop24} shows the cropped SR-2 region extracted from the spectral index map (left) and, for illustration, the corresponding zeroth-moment map integrated over the full velocity range (right). The zeroth-moment map is included only as a representative visualisation. For the actual analysis, the full CO image cube was first cropped to the SR-2 region, and all subsequent processing steps were carried out on this cropped cube.     
\begin{figure}
    \centering
    \includegraphics[width=0.95\linewidth]{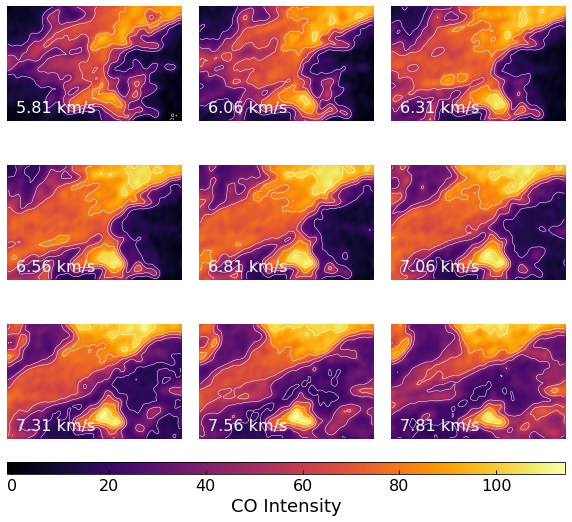}
    \caption{$^{12}$CO channel map from \protect\cite{2018ApJS..236...25K} for SR-2 region around 6.81 km s$^{-1}$ component.}
    \label{fig:co_chan}
\end{figure}

The channel map in Figure \ref{fig:co_chan} illustrates the emission around the velocity channel that shows the strongest correlation in SR-2 region. At velocities higher than this central channel, two spatially separated components are evident. Near the central velocity, the two structures appear to interact, while in the lower-velocity channel, they seem to merge into a single feature. One possible interpretation is that the higher-velocity component is moving into, or compressing, the lower-velocity component. However, without an independent distance estimate to these structures, this scenario cannot be confirmed.
  
\begin{figure}
    \centering
    \includegraphics[width=0.45\linewidth]{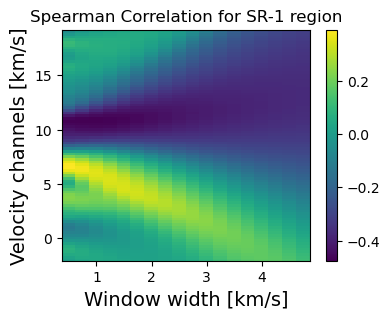}
    \includegraphics[width=0.45\linewidth]{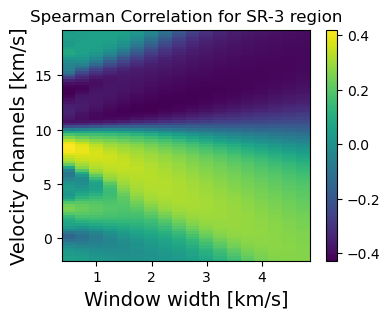}
    \includegraphics[width=0.45\linewidth]{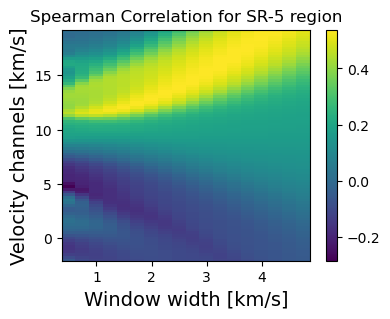}
    \includegraphics[width=0.45\linewidth]{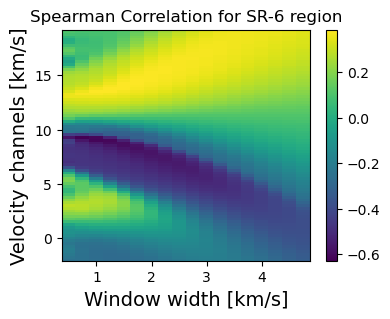}

    \caption{Same as Figure~\ref{fig: corr-plt} for SR-1, SR-3, SR-5, SR-6 regions.}
    \label{fig:corr_all}
\end{figure}

Figure \ref{fig:corr_all} shows the correlation results for SR-1, 3, 5, and 6. Among these regions, SR-1, SR-3, and SR-6 exhibit moderate to strong correlation, whereas SR-5 shows weaker correlation in comparison. 
\begin{figure}
    \centering
    \includegraphics[width=1\linewidth]{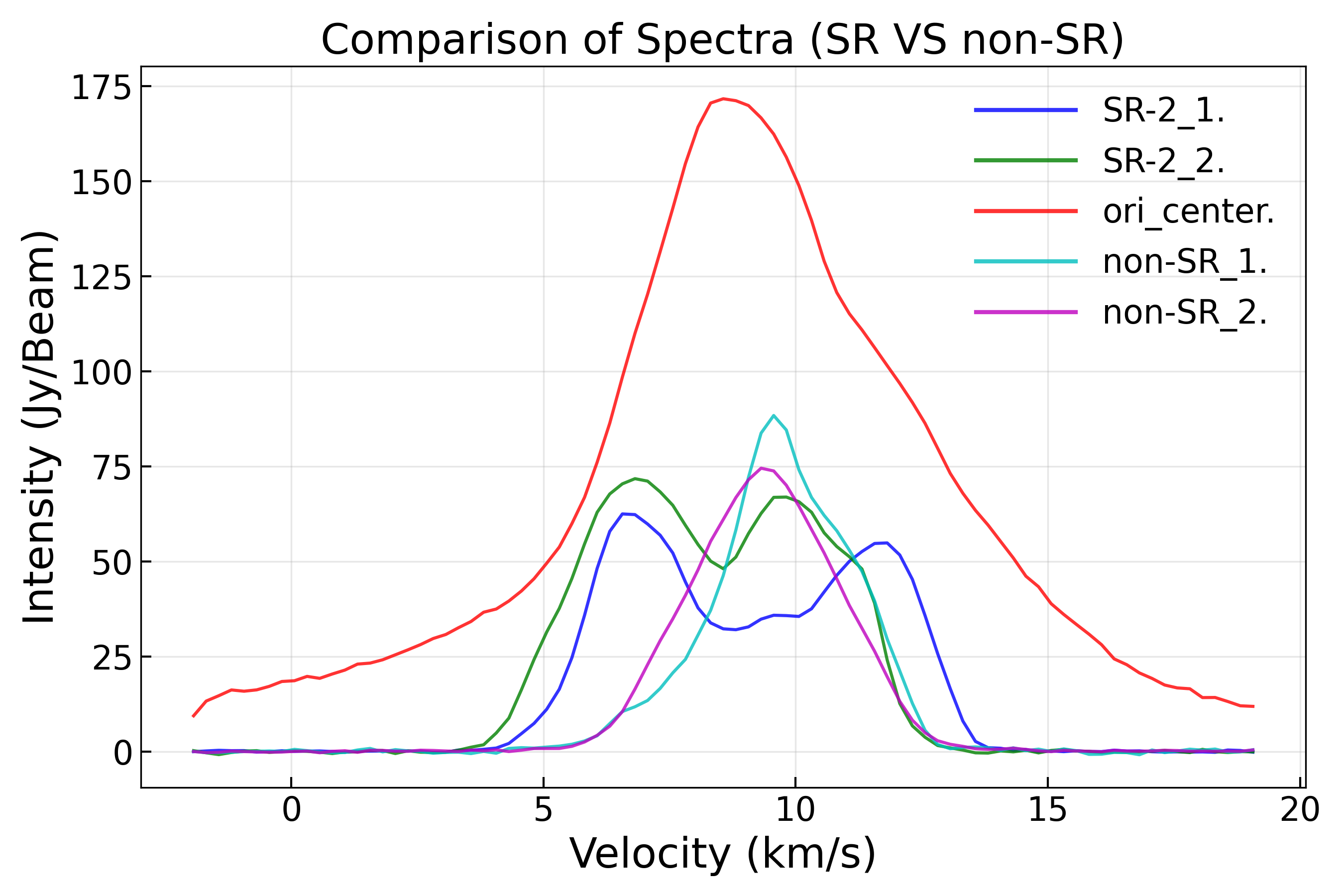}
    \caption{$^{12}$CO spectra from SR-2 region (blue and green), centre of the Orion Nebula (red) and away from the centre and SR regions (magenta and sky blue).}
    \label{fig:co-spectra}
\end{figure}

Figure \ref{fig:co-spectra} shows the $^{12}$CO spectra extracted from different positions in the Orion Nebula. The red spectrum corresponds to the central part of the Integral Shaped Filament (ISF). The green and blue spectra are taken from two distinct locations within the SR-2 region. The purple and cyan spectra are extracted from positions that do not belong to the central ISF or to any of the SR subregions.

\bsp	
\label{lastpage}
\end{document}